\pdfoutput=1
\documentclass[fleqn,usenatbib]{mnras}

\usepackage{newtxtext,newtxmath}

\usepackage[T1]{fontenc}
\usepackage{ae,aecompl}
\usepackage{times}

\usepackage{graphicx}	
\usepackage{amsmath}	
\usepackage{amssymb}	

\newcommand{\bs}{\boldsymbol}

\newcommand{\percent}{\,\mathrm{per cent}}

\title[
Kinematics of the Galactic bar-bulge 
]{Transverse kinematics of the Galactic bar-bulge from VVV and Gaia
}

\author[J. L. Sanders et al.]{
Jason L. Sanders,$^{1}$\thanks{E-mail: jls@cam.ac.uk (JLS), nwe@ast.cam.ac.uk (NWE)}
Leigh Smith,$^{1,2}$
N. Wyn Evans,$^{1}$
Philip Lucas$^{2}$
\\
$^{1}$Institute of Astronomy, University of Cambridge, Madingley Rd, Cambridge, CB3 0HA, UK\\
$^{2}$School of Physics, Astronomy and Mathematics, University of Hertfordshire, College Lane, Hatfield AL10 9AB, UK\\
}

\date{Accepted XXX. Received YYY; in original form ZZZ}

\pubyear{2019}

\begin{document}
\label{firstpage}
\pagerange{\pageref{firstpage}--\pageref{lastpage}}
\maketitle

\begin{abstract}
We analyse the kinematics of the Galactic bar-bulge using proper motions from the ESO public survey Vista Variables in the Via Lactea (VVV) and the second Gaia data release. Gaia has provided some of the first absolute proper motions within the bulge and the near-infrared VVV multi-epoch catalogue complements Gaia in highly-extincted low-latitude regions. We discuss the relative-to-absolute calibration of the VVV proper motions using Gaia. Along lines of sight spanning $-10<\ell/\,\mathrm{deg}<10$ and $-10<b/\,\mathrm{deg}<5$, we probabilistically model the density and velocity distributions as a function of distance of $\sim45$ million stars. 
The transverse velocities confirm the rotation signature of the bar seen in spectroscopic surveys. The differential rotation between the double peaks of the magnitude distribution confirms the X-shaped nature of the bar-bulge. Both transverse velocity components increase smoothly along the near-side of the bar towards the Galactic centre, peak at the Galactic centre and decline on the far-side. The anisotropy is $\sigma_\ell/\sigma_b\approx1.1-1.3$ within the bulk of the bar, reducing to $0.9-1.1$ when rotational broadening is accounted for, and exhibits a clear X-shaped signature. The vertex deviation in $\ell$ and $b$ is significant $|\rho_{\ell b}|\lesssim0.2$, greater on the near-side of the bar and produces a quadrupole signature across the bulge indicating approximate radial alignment. We have re-constructed the 3D kinematics from the assumption of triaxiality, finding good agreement with spectroscopic survey results. In the co-rotating frame, we find evidence of bar-supporting x1 orbits and tangential bias in the in-plane dispersion field.
\end{abstract}

\begin{keywords}
Galaxy: bulge -- kinematics and dynamics -- structure -- centre
\end{keywords}



\section{Introduction}
The Milky Way bulge is the only Galactic bulge which we can map in full kinematic detail. The combination of photometric, spectroscopic and proper motion studies admits the detailed study of individual stellar populations within the Galactic bulge, revealing the formation mechanism and subsequent evolution of this component. The favoured theoretical picture is of a dynamically-formed bar-bulge that first forms from bar instabilities in the disc before buckling and vertically spreading into the observed bulge component. A classical bulge component formed from early accretion may also be present, though this remains controversial~\citep{Shen10,DiMatteo15}.

Large-scale photometric studies (Optical Gravitational Lensing Experiment [OGLE], Two Micron All-Sky Survey [2MASS], UKIRT Infrared Deep Sky Survey [UKIDSS], Vista Variables in the Via Lactea [VVV]) of the red giants towards the Galactic centre have produced a coherent picture of the bulge as an elongated triaxial bar structure viewed near end-on (major axis at $\sim30\,\mathrm{deg}$ to the Galactic centre line-of-sight) \citep{Stanek1997,Saito2011,WeggGerhard2013,Simion2017}. Beyond $|\ell|\approx 10\,\mathrm{deg}$, the bar-bulge gives way to the \emph{long bar}, which has been traced out to $\ell\sim40\,\mathrm{deg}$ \citep[$\sim 5.5\,\mathrm{kpc}$, ][]{Wegg2015} and appears to be continuously connected to the bar-bulge. This suggests both components are dynamically-linked and co-rotate though this has not been demonstrated conclusively.

Going beyond its structural properties, the dynamical structure of the bar-bulge has been most clearly elucidated by spectroscopic studies. The line-of-sight mean velocities from the Bulge Radial Velocity Assay \citep[BRAVA,][]{BRAVA}, Abundances and Radial velocity Galactic Origins Survey \citep[ARGOS,][]{ARGOS}, Apache Point Observatory Galactic Evolution Experiment \citep[APOGEE,][]{APOGEE,APOGEEDR14}, Giraffe Inner Bulge Survey \citep[GIBS,][]{GIBS} and the Gaia-ESO survey \citep{GaiaESO} all demonstrate the cylindrical rotation expected for a dynamically-formed bulge~\citep{Howard09}. Furthermore, dissection of the populations by spectroscopic metallicity suggest the presence of a classical bulge component for metal-poor populations \citep{Ness13}, whilst the metal-rich population is characterised by orbits typical of buckled bars~\cite[e.g.,][]{Williams16}. The photometric and spectroscopic measurements have been successfully modelled by \cite{Portail2017} who inferred a pattern speed of $\Omega_\mathrm{p}=(39\pm3.5)\mathrm{km\,s^{-1}\,kpc^{-1}}$ placing corotation near $\sim 6\,\mathrm{kpc}$ consistent with the observations of the long bar.

With the arrival of data from the Gaia satellite \citep{Gaia2018}, there is the opportunity to complement the spectroscopic studies of the bar-bulge with large-scale proper motion surveys to further pin down its dynamics and formation process. Traditionally, proper motion studies require a set of background sources with assumed zero proper motion (e.g. quasars) to anchor the proper motion zero-point. In the bulge region, high extinction and high source density means background reference sources are hard to come by and studies have been restricted to relative proper motions, for which primarily dispersions have been measured. The earliest proper motion study of bulge stars was undertaken by \cite{Spaenhauer1992}, who extracted $\sim400$ K and M giants from photographic plates in Baade's window, $(\ell,b)=(1.02,-3.93)\,\mathrm{deg}$ finding $(\sigma_\ell,\sigma_b)\approx(115,100)\,\mathrm{km\,s}^{-1}$. Further ground-based studies have focussed primarily on giant stars in other windows \citep[Plaut's window and NGC 6558,][]{Mendez1996,Vieira2007,Vasquez2013}. The OGLE survey opened the possibility of measuring ground-based proper motions over 45 bulge fields \citep{Sumi2004,Rattenbury2007,Poleski2013} distributed along $b\approx-3.5\,\mathrm{deg}$ and the minor axis. \cite{Rattenbury2007} quantified for the first time variation in the proper motions with Galactic coordinates finding both $\sigma_\ell$ and $\sigma_b$ increasing towards the Galactic centre. Space-based proper motions have been measured using the Hubble Space Telescope (HST) for select (low extinction) fields including Baade's window \citep{Kuijken2002}, Sgr I \citep{Kuijken2002,Clarkson2008,Clarkson2018}, NGC 6553 \citep{Zoccali2001}, NGC 6528 \citep{Feltzing2002} and three minor axis fields \citep{Soto2014}. The largest HST survey was conducted by \cite{Kozlowski2007}, who measured proper motions of main sequence stars in $35$ fields distributed around Baade's window. The gradients of \cite{Rattenbury2007} were not clearly reproduced by this study although the reported uncertainties were a factor of a few larger. To date, the coverage of the bulge by proper motion studies is sparse and, besides the study of \cite{Rattenbury2007}, variation of dispersion within the bulge has not been conclusively demonstrated requiring comparison between different studies.

A number of studies have combined spectroscopy with proper motion surveys to reveal full 3D kinematics of the bulge. For instance, \cite{Zhao1994} combined the results of \cite{Spaenhauer1992} with radial velocity data to measure the vertex deviation of the bulge confirming its triaxiality. Further studies with spectroscopic data have elaborated on this result \citep{Ha00,Soto2007} and demonstrated the variation of kinematics with metallicity \citep{Babusiaux2010,Hill2011,Vasquez2013} revealing the relative contributions of the bar-bulge and a classical bulge (although for only limited fields).

The new astrometric data from the Gaia satellite \citep{Gaia2018} opens the possibility of fully characterizing the transverse velocity field of the bar-bulge. The second Gaia data release provided proper motions for $1.3$ billion stars with $G\lesssim21$ across the whole sky and so extends the limited view of previous proper motion surveys. Red clump stars at the Galactic centre have $G\approx16$, so Gaia is of limited use for highly extincted fields. However, the recent VIRAC catalogue \citep{VIRAC} for stars in the near-infrared $K_s$ band VVV catalogue extends the depth to which proper motions are available in high extinction regions (as $A_G/A_{Ks}\approx15$). As with previous proper motion studies of the bulge, the VIRAC catalogue produced by \citet{VIRAC} provides \emph{relative} proper motions i.e. the proper motions are not tied to an absolute reference frame. The Gaia second data release solves this issue, so for the first time absolute proper motions are available for bulge stars. In this paper, we briefly describe how absolute proper motions are computed by using bright stars in common between Gaia and VIRAC. Armed with this new proper motion catalogue, we present the transverse velocity structure of the bar-bulge. We decompose the density and velocity moments along the line-of-sight for fields $-10<b/\,\mathrm{deg}<5$ and $-10<\ell/\,\mathrm{deg}<10$.

The paper is laid out as follows. Section~\ref{Section::Data} describes the absolute proper motion catalogue created from Gaia and VIRAC, and the subset of data used in this study. We describe the methods employed in Section~\ref{Section::Method} focussing on extinction and completeness correction and the kinematic modelling employed. In Section~\ref{Section::Results}, we present the results of our analysis, before extracting the full 3d kinematics under the assumption of triaxiality in Section~\ref{Section::TriaxialStructure}.
We close with our conclusions in Section~\ref{Section::Conclusions}. In a companion paper (Sanders, Smith \& Evans, submitted, Paper II), we use our results to estimate the pattern speed of the bar using the continuity equation.

\section{Data}\label{Section::Data}

\subsection{Astrometry}
The VISTA Variables in the Via Lactea (VVV) survey \citep{Minniti2010,Saito2012} provides two epochs of $ZYJH$ imaging and many more epochs of $K_s$ band imaging over 5 years from the VISTA Infrared Camera (VIRCAM), covering $560$ square degrees of the southern Galactic plane and bulge.
The VVV Infrared Astrometric Catalogue version 1 \citep[VIRAC, ][]{VIRAC} is a proper motion catalogue for $\sim 300$ million sources derived from VVV survey data. VIRAC v1 uses up to several hundred epochs of $K_s$ band data per source by combining the overlapping VIRCAM observations (pawprint sets) necessary to obtain continuous coverage over the VIRCAM $1.65$ square degree field of view.
VIRAC v1 astrometric accuracy varies across the survey due to varying observing cadence and source density, but typical errors are $0.67\,\mathrm{mas\,yr}^{-1}$ for $11<K_s<14$, increasing to a few $\mathrm{mas\,yr}^{-1}$ at $K_s=16$.
One important caveat of VIRAC v1 proper motions is that they are relative to the mean motion of the local astrometric reference sources used, limiting its usefulness for studying kinematics over large scales.

\begin{figure}
    \centering
    \includegraphics[width=\columnwidth]{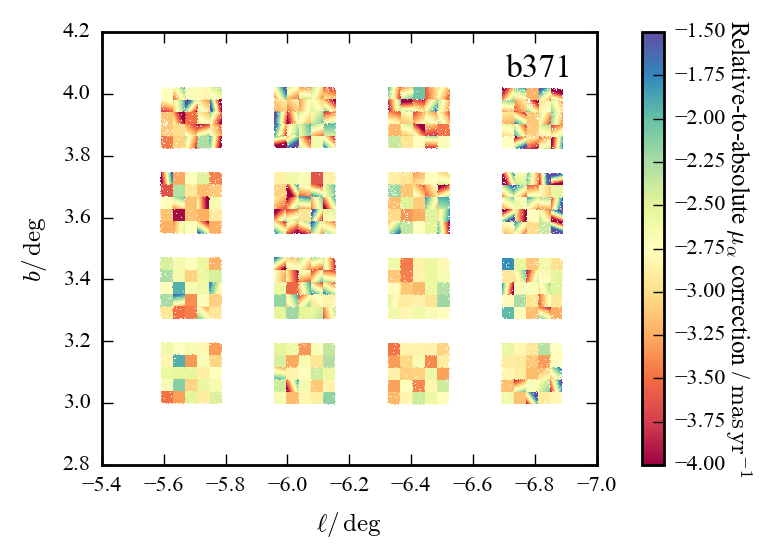}
    \caption{Relative to absolute proper motion corrections in $\alpha{}\cos{\delta}$ for sources in one pawprint set of VVV tile b371 (which contains $16$ arrays). The sub-array pattern is visible, as are regions in which linear (a gradient across the sub-array) or constant offset (flat colour in the subarray) correction methods were applied depending on VVV source density and the number of available Gaia counterparts.}
    \label{fig:rel2abs}
\end{figure}

With the release of the Gaia DR2 data \citep{Gaia2016,Gaia2018}, it is now possible to tie the relative proper motions of VIRAC v1 to an absolute reference frame using stars common to both catalogues.
We begin this process with VIRAC v1 intermediate data, proper motions generated from astrometric fits inside sub-arrays (each array is divided into $5\times{}5=25$ sub-arrays each covering $2.3\time2.3\mathrm{arcmin}^2$), as these are free from the reference-frame distortions introduced by the averaging proper motion solutions across overlapping pawprint sets. A detailed description of the production of this intermediate data is provided in section 3 of \citet{VIRAC}.
For each sub-array we used one of three relative to absolute correction methods depending on the number of available Gaia reference sources and the VVV source density. For relatively sparse sub-arrays with sufficient Gaia reference sources, we fit and apply a 6 coefficient linear function describing proper motion reference frame shift, skew and magnification as a function of sub-array position. For more dense sub-arrays with relatively few Gaia reference sources, we simply measure the average offsets between VIRAC and Gaia proper motions in both dimensions and apply these to the VIRAC proper motions. For sub-arrays with very few ($<10$) available Gaia reference sources, we revert to a 6 coefficient linear solution as described above but using Gaia reference sources across the entire array.
Potential reference sources are selected from Gaia DR2 as those with 5 parameter solutions, \textit{astrometric\_gof\_al} $<3$, and \textit{astrometric\_excess\_noise\_sig} $<2$, and from VIRAC as having no proper motion error flags (see section 4.2 of \citealt{VIRAC}). \cite{Lindegren2018} also discuss using the unit weight error for Gaia DR2 astrometric quality cuts, and since starting this work the reduced unit weight error has been officially recommended as the astrometric quality indicator. The pools of potential reference sources are matched within a 1\arcsec{} radius keeping only the best matches for each source.
Figure \ref{fig:rel2abs} shows relative to absolute proper motion corrections for sources of one pawprint set of VVV tile b371, the sub-array divisions are visible, as are regions in which the linear and constant offset correction methods are applied.

In all cases, we add the uncertainty on the relative to absolute correction in quadrature to the VIRAC v1 relative proper motion uncertainty to produce the uncertainty on the absolute proper motions. This procedure naturally transfers any Gaia DR2 systematic issues to the VIRAC catalogue, but typically the magnitude of known Gaia DR2 systematic issues is much smaller than the random errors on the VIRAC proper motions. 

Once this process of relative to absolute proper motion correction is performed, we verify that proper motion measurements of the same sources from overlapping pawprint sets (which are essentially independent measurements) were consistent within their uncertainties and then average these measurements following the procedure described in section 4.3 of \citet{VIRAC}. The resulting catalogue of absolute proper motions is dubbed VIRAC v1.1.

\subsection{Photometry}

We primarily work with $K_s$ photometry provided in the VIRAC catalogue (processed using the CASU pipelines) and supplement with $J$ when building an extinction map and for data selection. We also use the corresponding uncertainties $\sigma_i$ for quality cuts. We check the calibration of the photometry against the 2MASS catalogue \citep{Skrutskie2006} using the relationships reported by \cite{GonzalezFernandez2018}: 
$J_2=J+0.0703(J-K_s)$ and $K_{s2}$=$K_{s}-0.0108(J-K_s)$ (CASU v1.3). These expressions ignore extinction which produces corrections of $-0.003\mathrm{E}(J-K)$ in $J$ and $0.001\mathrm{E}(J-K)$ in $K$ \citep{GonzalezFernandez2018} so are negligible for all but the most highly extincted stars. For VIRAC catalogue entries with $11.5<J<14$, $11<K_s<14$ and $\sigma_J,\sigma_{Ks}<0.2$, we select the nearest 2MASS source within $1\,\mathrm{arcsec}$ with \texttt{ph\_qual} = A,B,C,D and \texttt{cc\_flg} = 0 in both $J_2$ and $K_{s2}$. For $0.5\,\mathrm{deg}$ by $0.5\,\mathrm{deg}$ fields, we compute the median offset between the VVV bands transformed to the 2MASS system and the 2MASS bands. We find that the difference in $K_s$ band varies by $\lesssim0.02\,\mathrm{mag}$ across the bulge region (corresponding to a distance systematic of $\sim70\,\mathrm{pc}$ at the Galactic centre) and $J$ by $\lesssim0.03\,\mathrm{mag}$ (corresponding to $K_s$ variations of $\sim0.014\,\mathrm{mag}$ for our assumed extinction law). The photometric systematic uncertainties are therefore negligible for our application.

\subsection{Selection}
Red clump giants have been used as a tracer of the structure of the bulge in numerous studies \citep{Stanek1997,Saito2011,WeggGerhard2013,Simion2017} due to their standard candle nature. They appear as a clear peak in bulge colour-magnitude diagrams, lying at $(J-K_s)_0\approx0.6$ and between $K_{s0}\approx12$ and $K_{s0}\approx14$ depending on Galactic longitude $\ell$ (subscript $0$ denotes unextincted -- we describe extinction correction in the following section).
We select data from VIRAC v1.1 and cross-match to the Gaia DR2 catalogue using a $1\,\mathrm{arcsec}$ radius, not accounting for the proper motions and epoch difference. From this combined catalogue, we select sources according to the following criteria:
\begin{itemize}
    \item $11.5<K_{s0}<14.5$,
    \item $0.4<(J-K_s)_0<1$ (if $J$ available),
    \item $\sigma_{Ks}<0.2$,
    \item $\varpi<0.75\,\mathrm{mas}$ or $\varpi/\sigma_\varpi<5$ (if $\varpi$ available),
\end{itemize}
where $\varpi$ is the Gaia parallax and $\sigma_\varpi$ its uncertainty. Furthermore, we remove stars within $3$ half-light radii of known globular clusters \citep[][2010 edition]{Harris1996}. The magnitude selection encompasses the bulge red clump peak whilst also providing sufficient stars at $11.5<K_{s0}<12$ and $14<K_{s0}<14.5$ to estimate the broader disc giant component over the range $12<K_{s0}<14$. At $K_s<11.5\,\mathrm{mag}$, non-linearity and saturation affect the VVV magnitudes \citep{Gonzalez2013}. The colour selection removes many nearby contaminant main sequence disc stars. However, if $J$ is unavailable, we still include the source in our selection so as not to affect the $K_s$ completeness \citep{WeggGerhard2013}. The parallax cut is a measure to remove nearby dwarf contaminants. 

\begin{figure}
$$\includegraphics[width=\columnwidth]{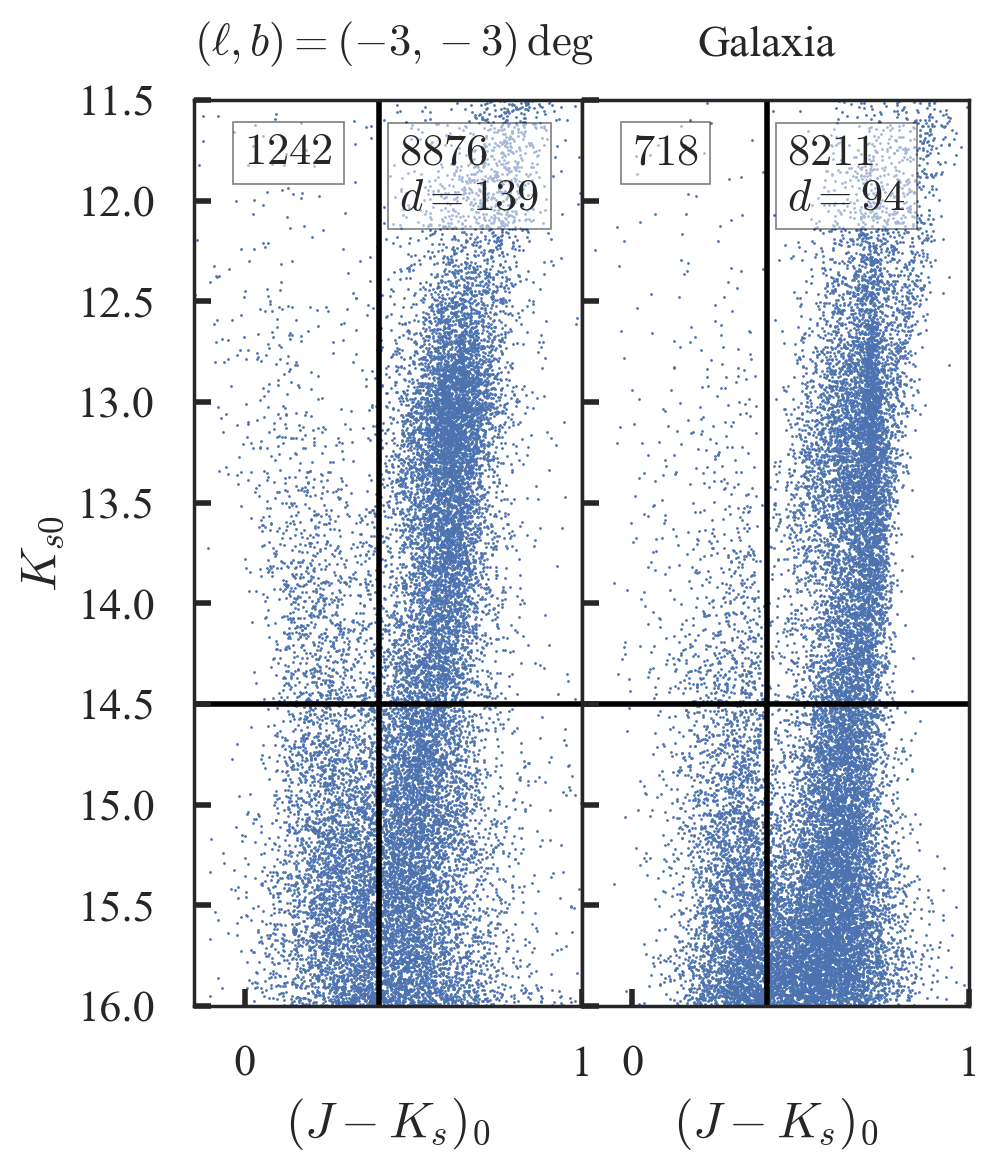}$$
\caption{Unextincted colour-magnitude diagrams for VVV (left) and Galaxia (right) for a field centred on $(\ell,b)=(-3,-3)\,\mathrm{deg}$. Our selection corresponds to the top right box. The insets give the number of stars in each box and $d$ is the number of dwarfs in the top right box.}
\label{fig::Selection}
\end{figure}

We have simulated our selection using Galaxia \citep[v0.7.2, ][]{Sharma2011} with the default set of parameters. In Fig.~\ref{fig::Selection}, we display the colour-magnitude diagrams from VVV and Galaxia for a $0.04\,\mathrm{deg}^2$ field at $(\ell,b)=(-3,-3)\,\mathrm{deg}$. Brighter than $K_{s0}=14.5\,\mathrm{deg}$, there are both giants primarily located in the centre of the Galaxy and foreground blue main sequence stars. The colour cut efficiently removes these. A caveat is that we select stars that don't have $J$ magnitudes. Some of these stars could be blue main sequence. However, for our test field the probability of this is at most $\sim1242/8876\approx14\percent$, but given a lack of $J$ measurement it is more likely to be a redder star so the probability is significantly lower. 

Within the giant selection box, there are nearby lower main sequence stars but these are subdominant. From Galaxia, we find $94/8211\approx1\percent$ and in the data the cut on parallax removes $139$ of $8876$. The similar ratios give us confidence we are removing most contaminating dwarfs with the parallax cut.
At higher latitudes, $b=-10\,\mathrm{deg}$ the dwarf contamination fraction increases to $\sim17\percent$ but checking with Galaxia, many of these dwarfs are within $\sim1.5\,\mathrm{kpc}$. Removing these reduces the dwarf contamination fraction to $5\percent$. Near the plane, fewer Gaia parallaxes are available but the dwarf contamination is less of an issue as we are overwhelmed by the distant giants.

From Galaxia, we find that approximately a third of the selected giant stars are from the `disc' populations i.e. they are drawn from the disc density profiles as opposed to the bulge profile. As the disc density profile is broad compared with the bulge profile, it produces a more featureless magnitude distribution. From purely photometric data, it is very difficult to separate these populations so from the perspective of our modelling both populations together comprise the bulge.

For sources observed by both Gaia and VVV, we combine the equatorial proper motions from VIRAC and Gaia DR2 using inverse variance weighting. This assumes the estimates are independent which is not completely true due to our absolute-to-relative correction procedure for the VIRAC proper motions. We transform the resulting proper motions to Galactic coordinates propagating the covariances. When modelling the proper motions, we adopt the further quality cuts:
\begin{itemize}
    \item $\sigma_{\mu i}<1.5\,\mathrm{mas\,yr}^{-1}$.
    \item $|\mu_{i}-\langle\mu_i\rangle|<3\Delta_{\mu i}$.
\end{itemize}
For component $i$, $\sigma_{\mu i}$ is the proper motion uncertainty, $\langle\mu_i\rangle$ the median proper motion in a given field and $\Delta_{\mu i}$ the dispersion computed using the 16th and 84th percentile. These two quality cuts remove spurious proper motions \citep{VIRAC} as well as those which are highly uncertain offering little constraining power. Although the proper motion error is a function of $K_s$, so cutting on proper motion error preferentially removes fainter stars, our modelling will constrain $p(\mu|K_s)$ so this isn't a concern.

\section{Methodology}\label{Section::Method}

\begin{figure*}
$$\includegraphics[width=\textwidth]{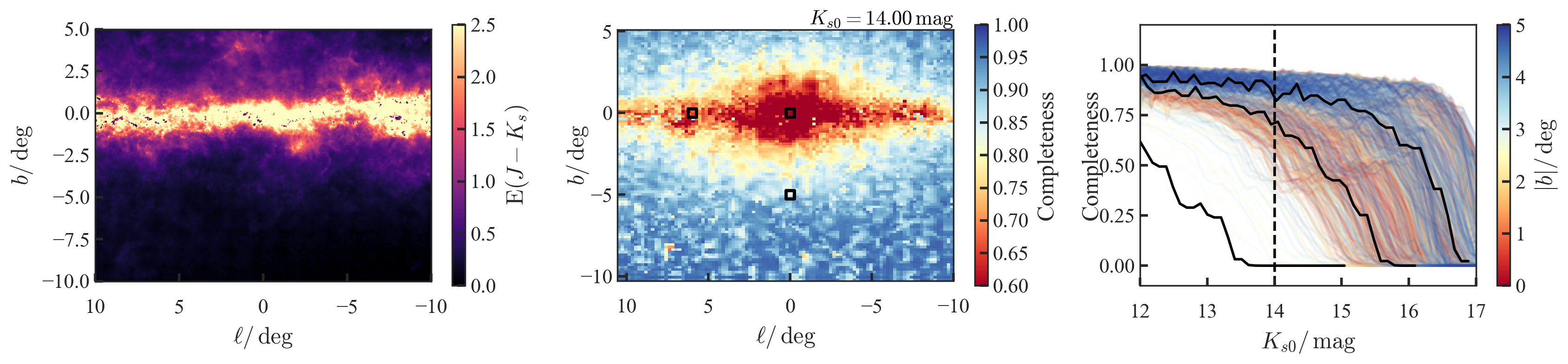}$$
\caption{Extinction and completeness: left panel shows the adopted extinction map, central panel the completeness (both source and proper motion completeness) at unextincted $K_{s0}=14\,\mathrm{mag}$ and the right panel shows the completeness at different lines of sight coloured by the Galactic latitude. The three black lines correspond to the square fields in the central panel.}
\label{ExtCompleteness}
\end{figure*}

Our aim is to deconvolve the volume density and velocity structure for the data described in Section~\ref{Section::Data}. We approach this problem in two stages: first, extracting the density structure in small on-sky bins by modelling the unextincted $K_s$ magnitude distribution for red giant stars, and secondly, combining the density structure with the proper motions to extract the transverse velocity distributions along the line-of-sight. Before embarking on this, we must model the extinction and completeness which are significant for many of the considered bulge fields. We then describe the entire kinematic model before explaining how it is broken into two stages.

\subsection{Extinction}
We follow \cite{Gonzalez2011} constructing a 2D extinction $\mathrm{E}(J-K_s)$ from the red clump giant stars (we ignore any extinction variation along the line-of-sight assuming the majority of the extinction is from foreground dust). We divide the VIRAC catalogue into fields of $3$, $5$ and $10\,\mathrm{arcmin}$ square for
$|b|<1.5\,\mathrm{deg}$, $1.5\,\mathrm{deg}<|b|<5\,\mathrm{deg}$ and $b<-5\,\mathrm{deg}$. In each field, we select stars in a diagonal colour-magnitude box to account for extinction such that $J<14+0.482(J-K_s-0.62)$ and $(J-K_s)>0.5$ and require uncertainties in $J$ and $K_s$ less than $0.2\,\mathrm{mag}$. In $(J-K_s)$, we find the peak $(J-K_s)_\mathrm{RC}$ using a multi-Gaussian fit and record the full-width half-maximum (FWHM) of this peak converting this into a standard deviation $\sigma(J-K_s)$. The unextincted red clump in Baade's window is
$(J-K_s)_{0,\mathrm{BW}}=0.62\,\mathrm{mag}$ \citep[obtained by transforming the \cite{Gonzalez2011} result to the VVV bands]{Simion2017} which we assume is valid across the bulge. In each field the extinction is given by $\mathrm{E}(J-K_s)=(J-K_s)_\mathrm{RC}-(J-K_s)_{0,\mathrm{BW}}$ with corresponding uncertainty $\sigma_{\mathrm{E}(J-K_s)}$ as \citep{WeggGerhard2013}
\begin{equation}
   \sigma_{\mathrm{E}(J-K_s)}^2 =\sigma(J-K_s)^2-\langle\sigma_J\rangle^2-\langle\sigma_{Ks}\rangle^2-\sigma(J-K_s)^2_\mathrm{RC},
\end{equation}
where we adopt an intrinsic red clump width of $\sigma(J-K_s)_\mathrm{RC}=0.05\,\mathrm{mag}$ and compute the median magnitude uncertainty $\langle\sigma_i\rangle$ in each field (if $\sigma_{\mathrm{E}(J-K_s)}^2<0$ we set $\sigma_{\mathrm{E}(J-K_s)}=0$). We use the \cite{Nishiyama2009} extinction law such that $A_{Ks}=0.482\mathrm{E}(J-K_s)$. This is similar to the extinction law $A_{Ks}=0.428\mathrm{E}(J-K_s)$ derived directly from VVV photometry by \cite{AlonsoGarcia2017} \citep[][also considered the \cite{Cardelli1989} extinction law and found the resulting large-scale bar-bulge properties to be unchanged]{WeggGerhard2013}. The results of our procedure are shown in the left panel of Fig.~\ref{ExtCompleteness}.

\subsection{Completeness}

There are two sources of incompleteness in our adopted catalogue. The first is incompleteness in the source catalogues used by VIRAC and the second is incompleteness due to each source not being assigned a proper motion. We assess the first of these using the method of \cite{Saito2012} and \cite{WeggGerhard2013} by inspecting the recovery of fake stars injected into the VVV images. For each bulge field, we choose the image with seeing closest to $0.75\,\mathrm{arcsec}$. For each array, we randomly add $5000$ stars with randomly selected magnitude $11<K_s<18\,\mathrm{mag}$ (using a Gaussian psf with FWHM of the seeing) and attempt to extract them using CASU's \texttt{imcore}. We repeat this procedure five times and record the average fraction extracted as a function of $K_s$. Naturally the completeness correlates with the source density, so is a strong function of $b$ with fields at $|b|\lesssim1\,\mathrm{deg}$ approximately $50\percent$ incomplete at $K_s=16\,\mathrm{mag}$.

Additionally, for each image we compare the true source catalogue to the sources with VIRAC proper motions and record the fraction with proper motions as a function of $K_s$. In general, this incompleteness is less severe than the source incompleteness. When analysing data we are concerned with the incompleteness as a function of unextincted magnitude. In Fig.~\ref{ExtCompleteness}, we display the total VIRAC completeness at $K_{s0}=14\,\mathrm{mag}$. We see how the completeness is a complex function of source density and extinction.

\subsection{Kinematic modelling}

From the described data, we wish to construct maps of the transverse velocity components $v_\ell$ and $v_b$ as a function of Galactic location within the Galactic bulge region. For a single sightline $(\ell,b)$, we construct $p(K_s,\bs{\mu}|\bs{\Sigma})$, where $K_s$ is the dereddened extinction $K_s$ (we drop the subscript zero from now on), $\bs{\mu}=(\mu_\ell,\mu_b)$ is the proper motion vector and
\begin{equation}
    \bs{\Sigma}=\begin{pmatrix}
    \sigma_{\mu\ell}^2&\rho_{\ell b}\sigma_{\mu\ell}\sigma_{\mu b}\\
    \rho_{\ell b}\sigma_{\mu\ell}\sigma_{\mu b}&\sigma_{\mu b}^2
    \end{pmatrix}
\end{equation}
is the uncertainty covariance matrix with $\sigma_{\mu i}$ the uncertainty in $\mu_i$ and $\rho_{\ell b}$ the correlation coefficient. We ignore uncertainties in $K_s$. This is given by
\begin{equation}
    p(K_s,\bs{\mu}|\bs{\Sigma}) = N^{-1}S(K_s)
    M(K_s,\bs{\mu}|\bs{\Sigma})
    ,
\end{equation}
where $M(K_s,\bs{\mu}|\bs{\Sigma})$ is the model for the bulge giant stars,
$S(K_s)$ is the completeness ratio and $N$ is a normalization constant given by
\begin{equation}
    N = \int_{11.5}^{14.5}\mathrm{d}K_s\,S(K_s)\int\mathrm{d}^2\boldsymbol{\mu}\,M(K_s,\bs{\mu}|\bs{\Sigma}).
    \label{equation::normalization}
\end{equation}
We write
\begin{equation}
    M(K_s,\bs{\mu}|\bs{\Sigma}) = \int \mathrm{d}s\, s^2 \rho(s) p(M_{Ks}) p(\bs{\mu}|M_{Ks},s,\bs{\Sigma}),
\label{eqn::fundamental}
\end{equation}
where $\rho(s)$ is the density profile as a function of distance $s$ along the line-of-sight specified by $(\ell,b)$. 
The giant branch luminosity function $p(M_{Ks})$ is evaluated at $M_{Ks}=K_s-2.171\ln s-10$ ($s$ in kpc).
The kinematic distribution $p(\bs{\mu}|M_{Ks},s,\bs{\Sigma})$ is in general a function of location $s$ and stellar type $M_{Ks}$. The kinematics are expected to be functions of both age and metallicity of the population. However, they will be weak functions of $M_{Ks}$ as along the giant branch we expect all stellar populations to contribute. Therefore, for simplicity we model $p(\bs{\mu}|s,\bs{\Sigma})$ only, for which we assume a Gaussian random field
\begin{equation}
    p(\bs{\mu}|s,\bs{\Sigma}) = \mathcal{N}(\bs{\mu}|\bs{m}(s),\bs{\Sigma}_m(s)+\bs{\Sigma}),
\end{equation}
where $\bs{m}(s)$ is the mean proper motion at each distance and $\bs{\Sigma}_m(s)$ the covariance\footnote{Throughout the paper we use the notation $\mathcal{N}(m,s)$ for a univariate normal distribution with mean $m$ and standard deviation $s$ and $\mathcal{N}(\bs{m},\bs{\Sigma})$ for a multivariate normal distribution with mean $\bs{m}$ and covariance $\bs{\Sigma}$.}. We model the density profile using a set of log-Gaussian basis functions.
\begin{equation}
\rho(s)=\sum_i^{N_c} w_i \mathcal{N}(\ln s|g_i,\sigma_i),
\end{equation}
where $g_i$ are the set of means in $\ln s$, $\sigma_i$ a set of widths, $\bs{w}$ is a simplex and $N_c$ the number of components which we set to three. 

\begin{figure*}
    \centering
    \includegraphics[width=\textwidth]{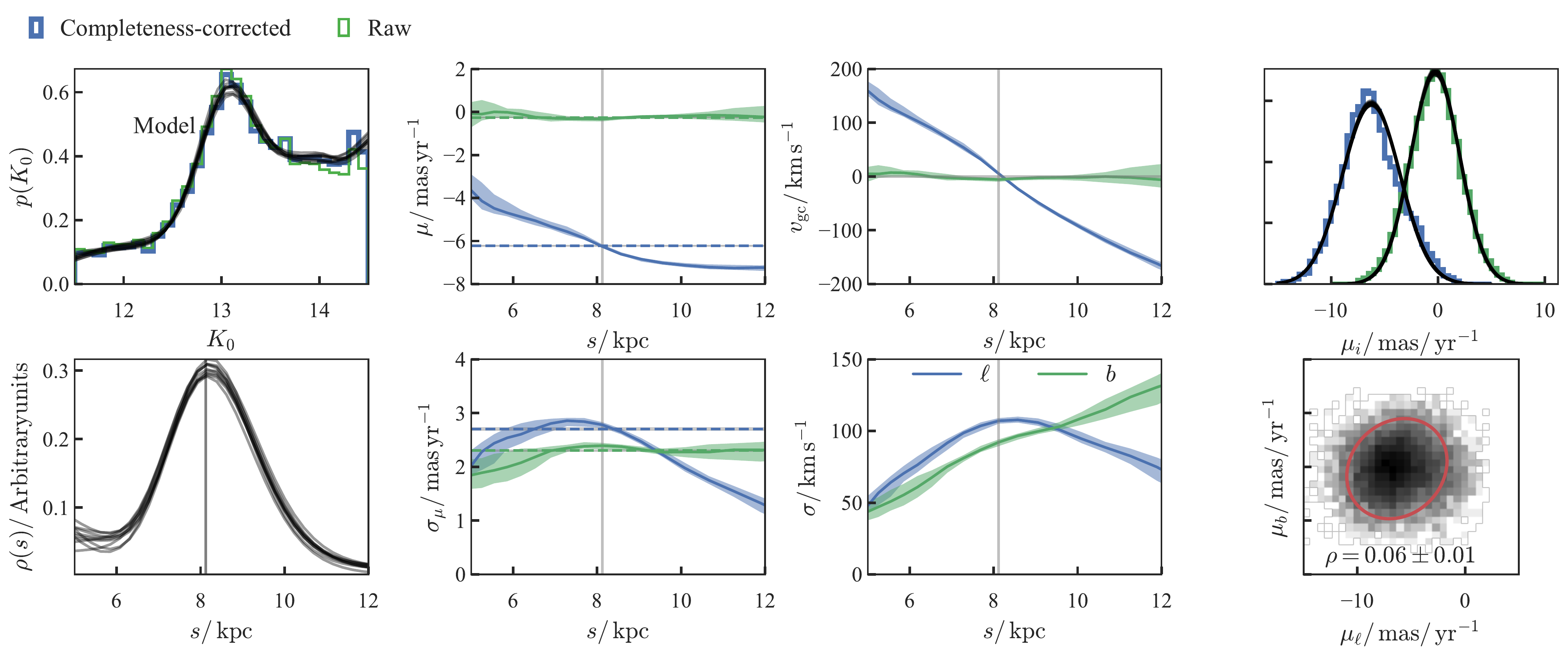}
    \caption{An example model fit for a $0.2\,\mathrm{deg}$ by $0.2\,\mathrm{deg}$ field centered on $(\ell,b)=(-3,-3)\,\mathrm{deg}$. The top-left panel shows the unextincted $K_s$ distribution (in green and completeness-corrected in blue). Samples from the model fit are shown in black. 
    The bottom left panel shows the giant distance distribution with the vertical line at the assumed Galactic centre distance $8.12\,\mathrm{kpc}$. The second column of panels show the run of the mean (top) and standard deviation (bottom) of the proper motions ($\ell$ in blue, $b$ in green) with 1$\sigma$ uncertainties. The dashed horizontal lines correspond to the distance-marginalized model. The third column are the corresponding space velocities corrected for the solar reflex.  The final column shows the proper motion distributions of the data: top panel histograms of proper motions in $\ell$ (blue) and $b$ (green) along with samples from the distance-marginalized model and bottom panel a 2d histogram (log-scaled) with a single contour from the multivariate Gaussian fit overplotted in red.
    }
    \label{fig:example_fit}
\end{figure*}
\subsection{Luminosity function}

Our model is for all giant stars towards the Galactic centre which can include both `disc' and `bulge' stars. In theory, we require different luminosity functions for each component. However, as the disc density profile is broad, details of its luminosity function (e.g. metallicity) are unimportant so we use a single luminosity function for both components.

The luminosity function $p(M_{Ks})$ is adopted from \cite{Simion2017} composed of three Gaussian peaks for the AGB, RGB and RC bumps along with a background RGB exponential: $M_{Ks}\propto\exp(a_g(M_{Ks}+1.53))$ where \cite{Simion2017} sets $a_g=a_{g0}=0.642$. \cite{Simion2017} allowed the mean magnitude of the red clump to vary in their modelling. Requiring a Galactic centre distance of $8\,\mathrm{kpc}$, \cite{Simion2017} found $M_{K,RC}=-1.63\,\mathrm{mag}$ consistent with the solar neighbourhood result of $M_{K,RC}=-1.61\,\mathrm{mag}$ \citep{Alves2000,Hawkins2017}. From stellar models \citep{GirardiSalaris,SalarisGirardi}, the red clump magnitude is a function of (at least) alpha-enhancement, age and metallicity. From different assumptions, \cite{WeggGerhard2013} employed a brighter red clump magnitude of $M_{K,RC}=-1.73\,\mathrm{mag}$ but also observed a significant vertical gradient in the inferred Galactic centre distance, probably due to metallicity gradients in the bulge \citep{Gonzalez2013}. Our early models also displayed a gradient in the inferred Galactic centre distance corresponding to a gradient in the absolute magnitude of the red clump of $\sim0.1\,\mathrm{mag\,kpc}^{-1}$. Fixing the distance to the Galactic centre as $R_0=(8.12\pm0.03)\,\mathrm{kpc}$ \citep{GravityCollab2018} gives $M_{K,RC}=-1.67\,\mathrm{mag}+0.1\,\mathrm{mag\,kpc}^{-1}|z/\,\mathrm{kpc}|$. This is pleasing as it implies the low latitude clump stars are on average super-solar metallicity, whilst those at higher latitudes are consistent with solar metallicity. For instance, in Baade's window $b=-4\,\mathrm{deg}$ we would expect $M_{K,RC}=-1.61$ as $z\approx0.57\,\mathrm{kpc}$, which is perfectly consistent with the solar neighbourhood \citep{Hawkins2017}. 

\subsection{Discretization}

We discretize both the integrals over distance and magnitude. We evaluate equation~\eqref{eqn::fundamental} by discretising the integral as
\begin{equation}
    M(K_s,\bs{\mu}) = \sum_i (\Delta\ln s)\, s_i^3 \rho(s_i) p(M_{Ks}(K_s,s_i)) p(\bs{\mu}|s_i),
\end{equation}
using a uniform grid in $\ln s$ of $N_s=24$ points between $4$ and $14\,\mathrm{kpc}$ (this upper limit choice only just encompasses the red clump at the far end of the bar -- tests with larger upper limit choices produce very similar results). For each grid-point, we model the proper motion distribution as a Gaussian
\begin{equation}
    p(\bs{\mu}|s_i,\bs{\Sigma}) = \mathcal{N}(\mu|\bs{m}_i,\bs{\Sigma}+\bs{\Sigma}_{mi}).
\end{equation}
Furthermore, we compute the normalization integral in equation~\eqref{equation::normalization} using a uniform grid of $N_K=30$ points in $K_s$ as
\begin{equation}
    N = \sum_j(\Delta K_s)\,S(K_{s,j})
    \sum_i(\Delta\ln s)\, s_i^3 \rho(s_i) p(M_{Ks}(K_{s,j},s_i)) 
    .
\end{equation}

\subsection{Inference}
We split our inference into two stages for each field. Each model is written in the probabilistic programming language, Stan \citep{Stan}. First, we evaluate the density profile $\rho(s)$ by running NUTS \citep{Hoffman2011} (for $1000$ iterations) with likelihood $p(K_s)=\int\mathrm{d}^2\bs{\mu}p(K_s,\bs{\mu})$ and priors
\begin{equation}
    \begin{split}
        &g_i\sim\mathcal{U}(4\,\mathrm{kpc},14\,\mathrm{kpc}),\\
        &a_g\sim\mathcal{N}(a_{g0},0.1a_{g0})\\
        &\ln(\sigma_i/\,\ln\mathrm{kpc})\sim\mathcal{N}(-1,2).
    \end{split}
\end{equation}
We infer the parameters $g_i$, $w_i$, $\sigma_i$ 
and $a_g$: that is the means, components and width of the Gaussian mixture for the distance distribution, 
and the slope of the giant branch absolute magnitude distribution. A further simplification is that we compute $p(K_s)$ on the grid and interpolate for each datum.

For each field, we take the median $\rho(s)$ fitted using this procedure and use the probability density function $p(\bs{\mu}|K_s)$ to infer the parameters $\bs{m}(s_i)$ and $\bs{\Sigma}(s_i)$. For speed reasons, we first infer the full proper motion covariance marginalized over distance\footnote{As we know the means and dispersions vary with distance, this procedure will be biased by the selection in $K_s$ (which is also affected by our error cut in proper motion) and is not representative of all stars in the range $11.5<K_s<14.5$ at each location. The average values obtained this way will vary with completeness, extinction etc. on the sky. However, we can check against our non-marginalized model results, doing the averaging along the line-of-sight in the correct way, and the results are very similar.} ($p(\bs{\mu}|s_i)=\mathcal{N}(\bs{\mu}|\bs{m},\bs{\Sigma}_m+\bs{\Sigma})$) and then fit the distance dependence of the two proper motion components, $\mu_\ell, \mu_b$ independently. In our later modelling, we assume that the intrinsic correlation between $\mu_\ell$ and $\mu_b$ is approximately independent of distance for computational speed, despite \cite{Clarkson2008} demonstrating that $\rho_{\ell b}$ varies with distance within the bulge. 
For each model, we run the NUTS sampler for $100$ iterations. We work with proper motions, $\bs{\mu}'$, shifted by the median and scaled by the standard deviation. We adopt a `smoothing spline' prior to regularize $p(\bs{\mu}'|s)$ as
\begin{equation}
    \begin{split}
        m'_{i}s_i&\sim\mathcal{N}(2m'_{i-1}s_{i-1}-m'_{i-2}s_{i-2},\tau_m(8\,\mathrm{kpc})),\\
        \sigma'_{i}s^{}_i&\sim\mathcal{N}(2\sigma'_{i-1}s^{}_{i-1}-\sigma'_{i-2}s^{}_{i-2},\tau_s(8\,\mathrm{kpc})),\>
    \end{split}
\end{equation}
where $m'_i$ and $\sigma'_i$ are the mean and variance for a single scaled and shifted proper motion component at distance $s_i$. These priors act to minimize the second derivatives with respect to log-distance of the mean and standard deviations of the physical velocities (hence multiplying by the distance). We further adopt the priors
\begin{equation}
    \begin{split}
        &m'_1\sim\mathcal{N}(0,5),\>
        m'_2\sim\mathcal{N}(0,5),\>
        \tau_{m}\sim\mathcal{N}(0,1),\>\tau_{m}>0,\\
        &\sigma'_1\sim\mathcal{N}(0,5),\>
        \sigma'_2\sim\mathcal{N}(0,5),\>
        \tau_{s}\sim\mathcal{N}(0,1),\>\tau_{s}>0.\\
    \end{split}
\end{equation}
For the distance-marginalized model, we adopt 
\begin{equation}
    \begin{split}
        m'\sim\mathcal{N}(0,5),\>
        \sigma'\sim\mathcal{N}(1,5),\>
        \rho_{\ell b}\sim\mathcal{N}(0,0.4),\>
    \end{split}
\end{equation}
where $\rho_{\ell b}$ is the covariance. In principle, we could also impose smoothing in the means and dispersions between different $(\ell,b)$ pixels. However, such a model would have a very large number of free parameters, so we analyse each field independently.

Transverse velocities are estimated in the heliocentric frame which we transform to the Galactocentric frame using a solar motion of $(U,V+V_c,W)=(11.1,245.5,7.25)\,\mathrm{km\,s}^{-1}$ \citep{Reid2004,GravityCollab2018,Schoenrich2010}.

\subsection{An example field}

In Figure~\ref{fig:example_fit}, we show the results of fitting the model to a $0.2\,\mathrm{deg}$ by $0.2\,\mathrm{deg}$ field at $(\ell,b)=(-3,-3)\,\mathrm{deg}$. This field has $9203$ stars satisfying our initial set of cuts and $8945/8952$ satisfying the proper motion cuts in $\ell/b$. The first column of panels shows the results of the density fit. Both the raw and the completeness-corrected unextincted magnitude distributions are displayed along with our model fit.
The peak of the bulge red giants is clearly visible at $K_0\sim13\,\mathrm{mag}$. This corresponds to the inferred density distribution in the lower panel peaking just beyond the assumed Galactic Centre distance. The second column shows the inferred run of the mean and standard deviation of the proper motions in the two components along with uncertainties. We also show the average inferred from marginalization over distance. The third column shows the corresponding solar-reflex-corrected velocities. The mean $\mu_b$ and $v_b$ are flat implying no net vertical flow at any position (as per expectation). The longitude motion is more interesting. The mean proper motion falls from $\sim-4\,\mathrm{mas\,yr}^{-1}$ at $s=5\,\mathrm{kpc}$ to $\sim-7\,\mathrm{mas\,yr}^{-1}$ beyond $s=10\,\mathrm{kpc}$. In the velocities, this corresponds to a `rotation curve' crossing zero near the Galactic centre distance and rising to $\sim150\,\mathrm{km\,s}^{-1}$ at the extremes. We also note the slight change in gradient $\sim1\,\mathrm{kpc}$ from the Galactic centre which mirrors the rotation curves derived from proper motion data by \cite{Clarkson2008}. 

The velocity dispersions rise from $(\sigma_\ell,\sigma_b)\approx(50,50)\,\mathrm{km\,s}^{-1}$ at $s=5\,\mathrm{kpc}$ towards the Galactic centre distance. $\sigma_\ell$ peaks at the Galactic centre at $\sim105\,\mathrm{km\,s}^{-1}$ and declines to $\sim70\,\mathrm{km\,s}^{-1}$ at $s=12\,\mathrm{kpc}$ whilst $\sigma_b$ continues to rise. The $\sigma_b$ behaviour is not exactly in agreement with our expectations. It appears that the model struggles to distinguish between stars at the Galactic centre and those beyond it, assigning a similar proper motion dispersion to both populations. The problem is not mirrored in $\ell$ possibly because the mean is evolving. We have inspected the $\mu_b$ distribution and it appears that for fainter magnitudes there is a slightly narrower peak embedded in a broader envelope, possibly highlighting deficiencies of simply modelling the distribution by a single Gaussian. However, this is not conclusive and needs further attention in future work. Although our selection is designed to remove disc contamination, nearby stars could also produce such a signal. Separation of the sample into two $(J-K_s)_0$ bins shows that the proper motion distributions (in both $\ell$ and $b$) only differ significantly for $K_s<12\,\mathrm{mag}$ meaning disc contamination is present but at low levels so is unlikely to cause such a signature. However, the dispersion at these locations could genuinely be high. At distances of $12\,\mathrm{kpc}$ we are observing Galactic heights of $600\,\mathrm{pc}$ corresponding to the classical thin-thick interface region.

Finally, Figure~\ref{fig:example_fit} also shows the proper motion distributions in each component and compare to our distance-marginalized model. 
We see that the $\mu_\ell$ distribution is approximately Gaussian but has a wing towards more positive $\mu_\ell$.
The $\mu_b$ distribution is highly symmetric. In the 2d histograms we see the correlation between the proper motions as well as the additional wing in $\mu_\ell$ corresponding to a slightly narrower range of $\mu_b$, which possibly corresponds preferentially to the `disc' population in the central Galaxy. The correlation is measured as $\rho_{\ell b}=0.06\pm0.01$ implying the velocity ellipsoid in $(\ell,b)$ is pointing towards to Galactic centre.

\section{Results}\label{Section::Results}

\begin{figure}
    \centering
    \includegraphics[width=\columnwidth]{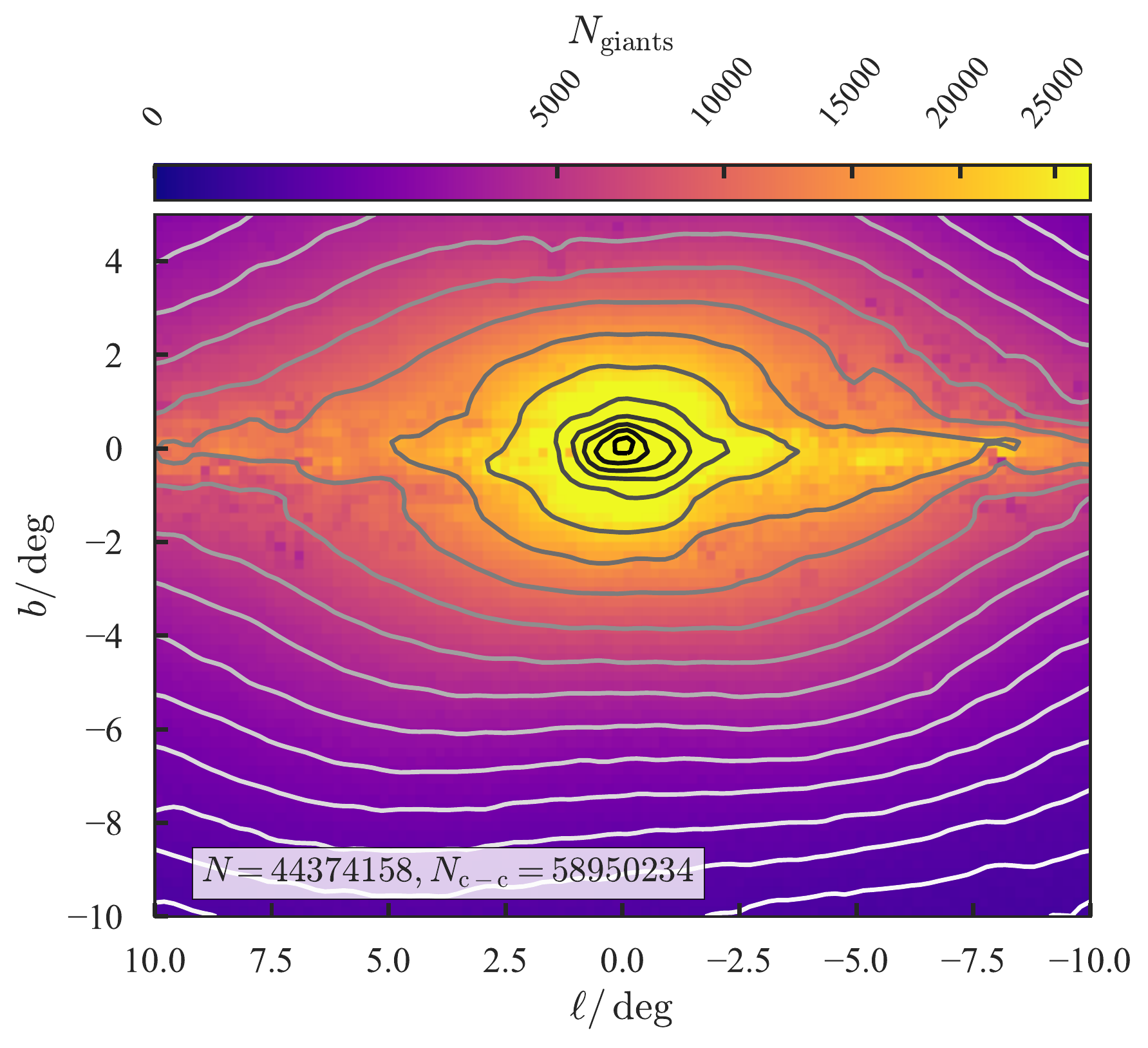}
    \caption{Completeness corrected bulge giant number density in bins of $0.2\,\mathrm{deg}$ by $0.2\,\mathrm{deg}$. The colourbar is square-root scaled and the contours are evenly spaced in $\log_{10}(N)$. The inset gives the total number of stars with ($N$) and without ($N_\mathrm{c-c}$) completeness correction.
    }
    \label{fig:number_counts}
\end{figure}
\begin{figure*}
    \centering
    \includegraphics[width=.87\textwidth]{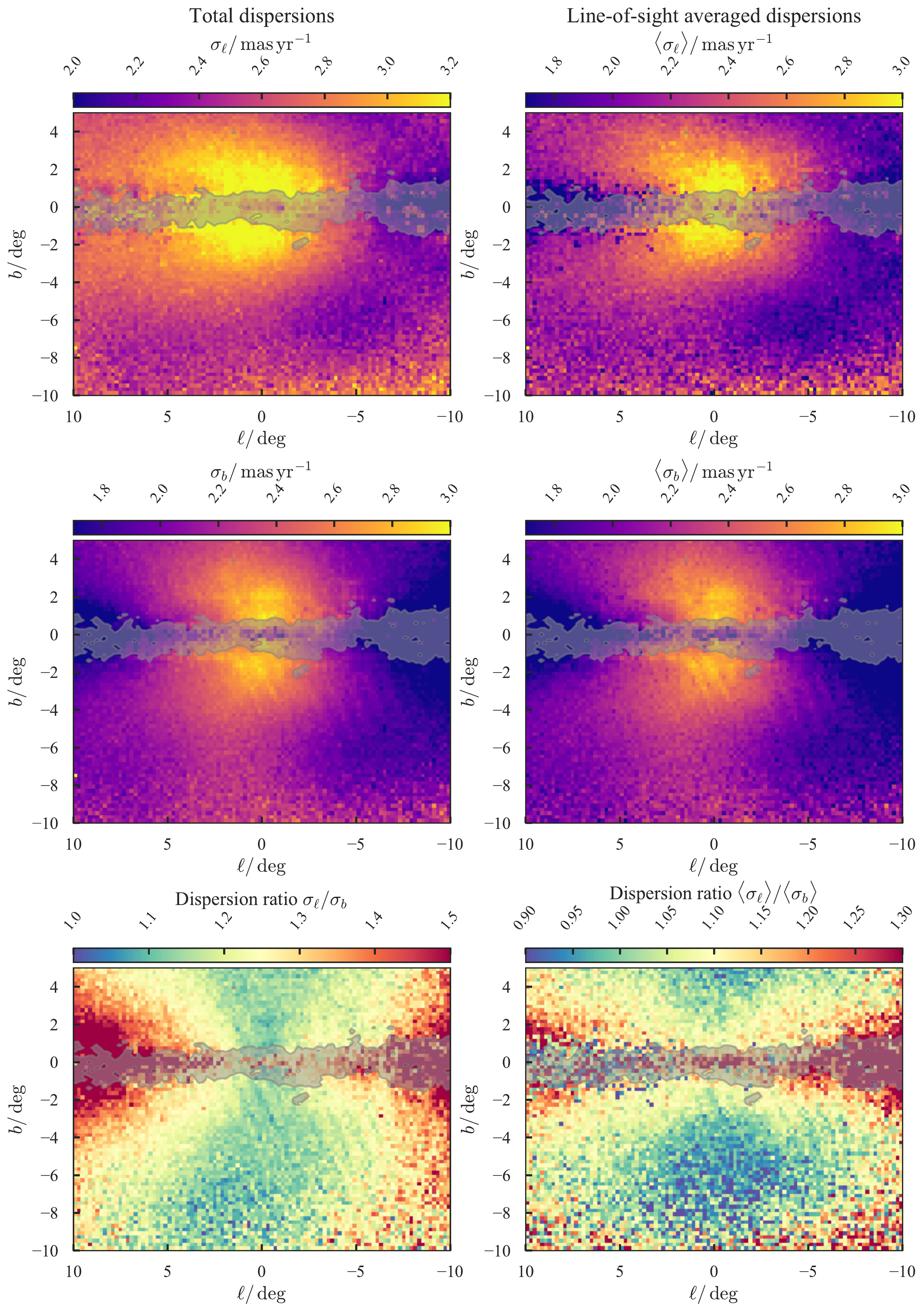}
    \caption{On-sky proper motion field: left panels show the dispersion of all giant stars within $11.5<K_s<14.5$ and right panels show the mean dispersion averaged along the line-of-sight of these stars. The latter of these removes the effects of rotational broadening.
    Top row shows the longitudinal proper motion, middle the latitudinal and bottom their ratio. Note the different scales, particularly for the top two and lower two plots.
    The grey overlay shows the region within which the extinction $A_{Ks}>0.8$.
    }
    \label{fig:allsky}
\end{figure*}

\begin{figure}
    \centering
    \includegraphics[width=\columnwidth]{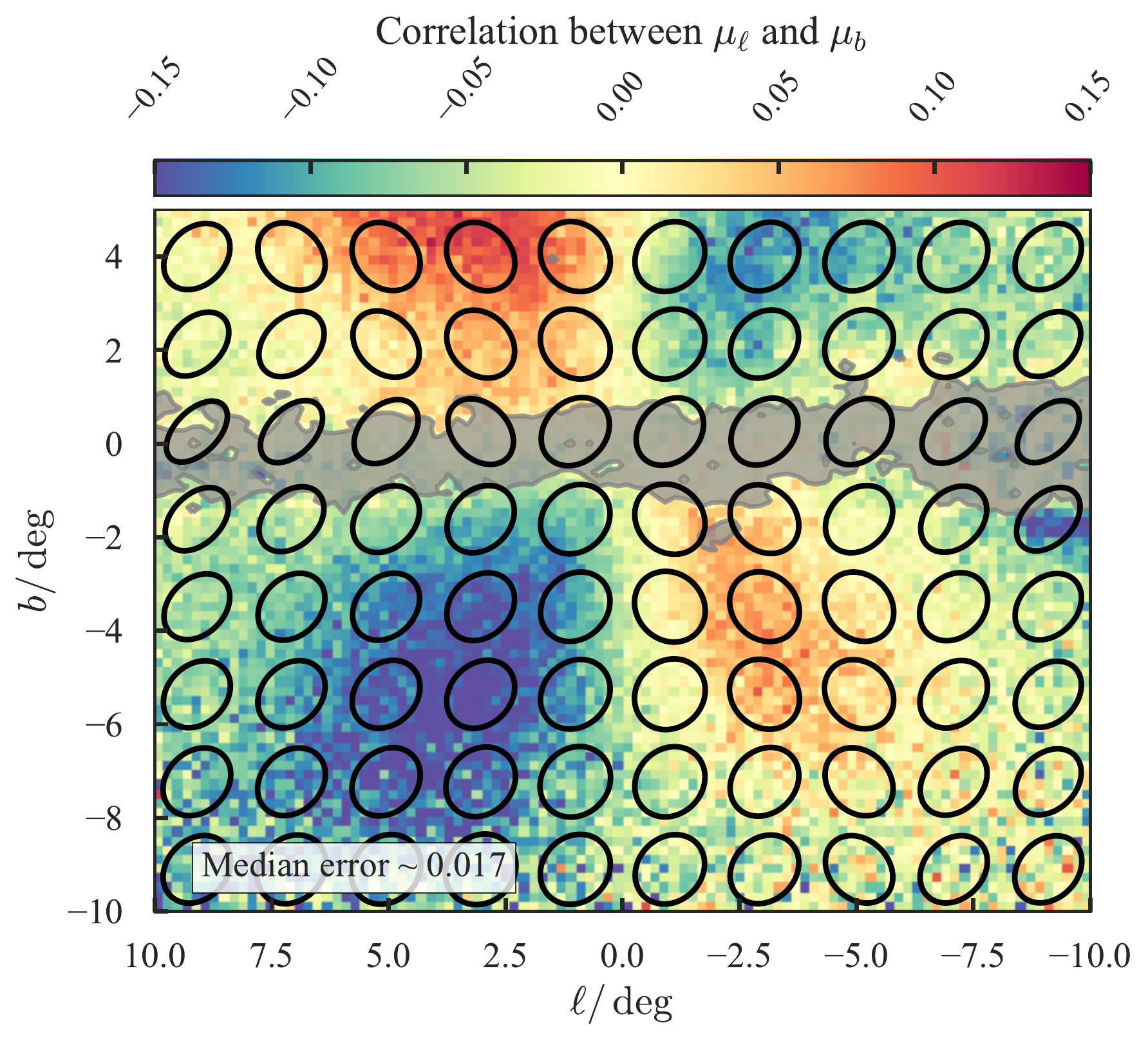}
    \caption{
    The correlation $\rho_{\ell b}$ between the two proper motion components. 
    The ellipses shows the shape of the velocity ellipsoid averaged over $9$ pixels. The ellipsoid is nearly everywhere radially-aligned. 
    The grey overlay shows the region within which the extinction $A_{Ks}>0.8$.
    }
    \label{fig:correlation}
\end{figure}

We have applied our modelling procedure to fields of $0.2\,\mathrm{deg}$ by $0.2\,\mathrm{deg}$ in $\ell$ and $b$ across the VVV bulge region ($-10<\ell/\,\mathrm{deg}<10$ and $-10<b/\,\mathrm{deg}<5$). This amounts to modelling $\sim44.5$ million stars
which when completeness corrected, represents $\sim59$ million stars. Of the $44.5$ million, approximately $24$ million are associated with the overdensity in the magnitude distribution and can reliably be attributed to the `bulge' component. The excess we attribute to a more extended `disc' background. However, our modelling considers all stars together, and makes no distinction between disc and bulge. We show the completeness-corrected number density in Fig.~\ref{fig:number_counts}. The bulge plus disc structure is clear with the asymmetric bar structure visible at high latitude.

\subsection{Distance-marginalized results}
We display the $\ell$ and $b$ proper motion dispersions for the bulge stars in the top panels of Fig.~\ref{fig:allsky}. The left column $\sigma_i$ shows the total dispersion of all stars with $11.5<K_s<14.5$ whilst the right column $\langle\sigma_i\rangle$ shows the dispersion at each distance averaged along the line of sight. The first of these is larger due to rotational broadening \citep{Zhao1994}, particularly in $\sigma_\ell$ -- the effect on $\sigma_b$ is not clearly distinguishable. The central regions of the bulge are kinematically hottest and both dispersions display a boxy-shaped profile similar to the density field. 
with a central collimated structure and a suggestion of an X-shape structure at larger $|\ell|$. The rotationally-broadened $\ell$ dispersion is boxier than its de-broadened counterpart. Both dispersions decline with increasing $|\ell|$ and $|b|$. In the plane, the dispersions are small probably reflecting the more dominant disc population in these bins and potentially issues with extinction modelling. We have overlaid in grey the region where the $A_{Ks}$ extinction is greater than $0.8\,\mathrm{mag}$, as this appears by-eye to be the region where fine structure due to extinction appears, suggesting our procedure is not valid here. Both dispersions display an asymmetry in $\ell$ with a tendency for larger dispersions at $\ell>0$ than $\ell<0$. This is consistent with viewing an edge-on bar with the nearer side at positive longitudes. At high latitude $|b|\gtrsim8\,\mathrm{deg}$, the dispersion increases slightly due to the dominance of the foreground disc and absence of the bulge in these fields.
We note that the $\ell$ and $b$ dispersions on the minor axis don't decline as strongly as off-axis. This is possibly due to the double-peaked nature of the minor-axis fields as we observe both sides of the X-shaped bulge.
There are clear artefacts in these maps: the vertical strips are due to the VVV imaging strategy and the diagonal features are due to the Gaia scanning law.

Inspecting the ratio of $\sigma_\ell$ to $\sigma_b$, we find the rotationally-broadened ratio is $\sim1.1$ on and around the minor axis and increases to $\sim1.2$ in an elongated X-shape for $|\ell|\gtrsim3\,\mathrm{deg}$. The low dispersion ratio of the disc is visible at low latitudes and large $|\ell|$. Removing the effects of rotational broadening, we see the dispersion ratio remains $\sim1.1$ at low latitudes whilst at higher latitudes, $b\sim-6\,\mathrm{deg}$ where the X-shape is contributing the ratio has dropped to $0.9-1$.

The correlation between the proper motion components is shown in Fig.~\ref{fig:correlation}. We see a clear quadrupole pattern corresponding to near alignment with the line towards the Galactic centre. The correlation is larger at positive longitude than negative longitude, possibly due to the geometric projection. The maximum amplitude of the correlation is around $\sim0.2$ at negative $\ell$ agreeing well with the more limited study of \cite{Kozlowski2007}. The typical uncertainty is around $0.02$. The correlation is small near the plane and increases in magnitude away from the plane reaching maximum values around $b\sim-5\,\mathrm{deg}$ for $\ell>0$ and $b\sim-4\,\mathrm{deg}$ for $\ell<0$ due to the triaxial shape. Beyond this, the number of bulge stars falls off so it is not clear whether radial alignment weakens or is diluted by the increased disc contribution. Along the minor axis, the ellipsoid is orientated $\sim90\,\mathrm{deg}$ with respect to the axis. Finally, we comment that the Gaia proper motions have star-to-star correlations on small scales which could affect our measurements. However, the signature we have extracted is sufficiently large-scale that it is insensitive to any systematics. 

\subsection{Distance-dependent results}

\begin{figure*}
    \centering
    \includegraphics[width=\textwidth]{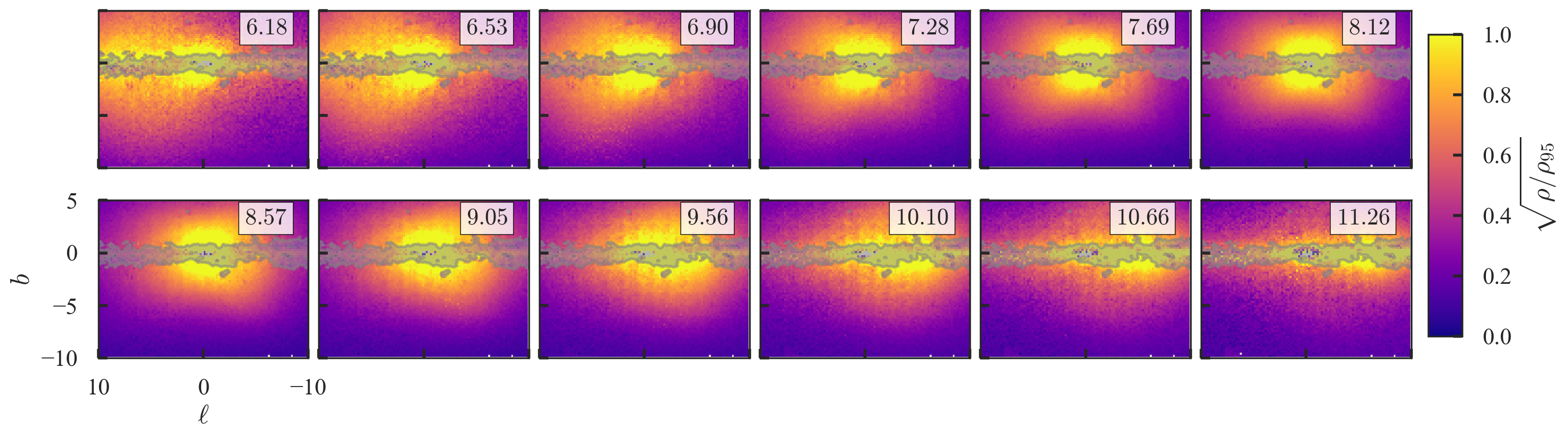}
    \includegraphics[width=\textwidth]{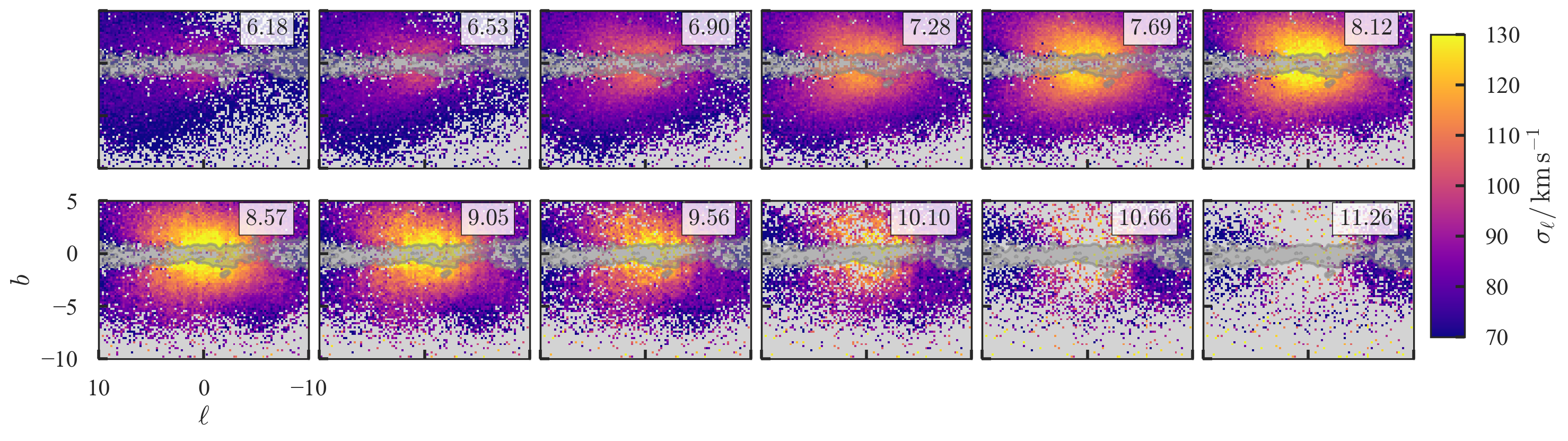}
    \includegraphics[width=\textwidth]{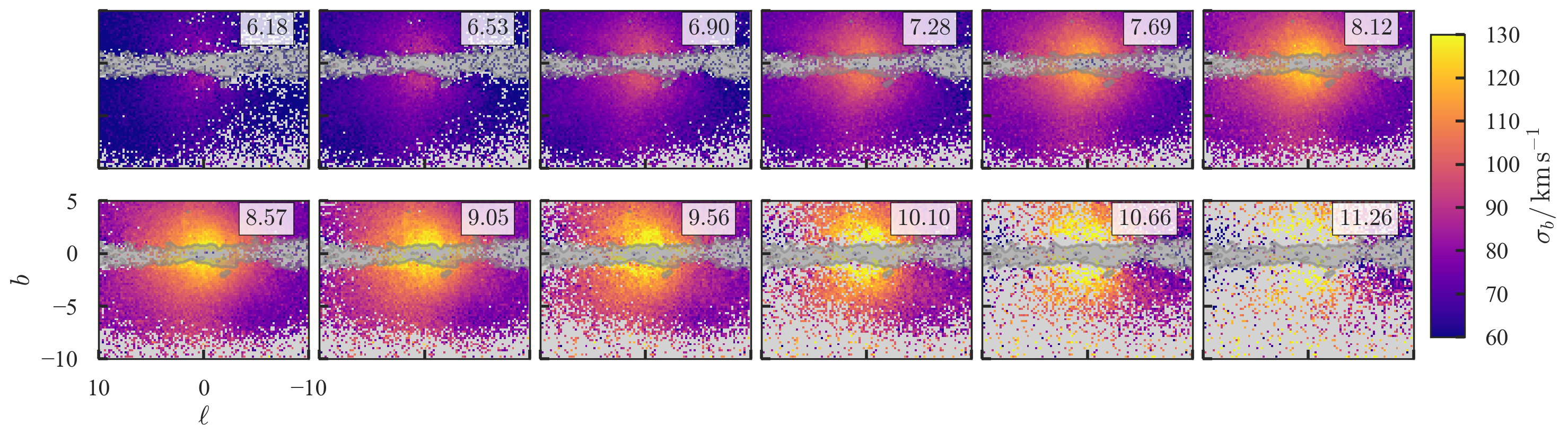}
    \includegraphics[width=\textwidth]{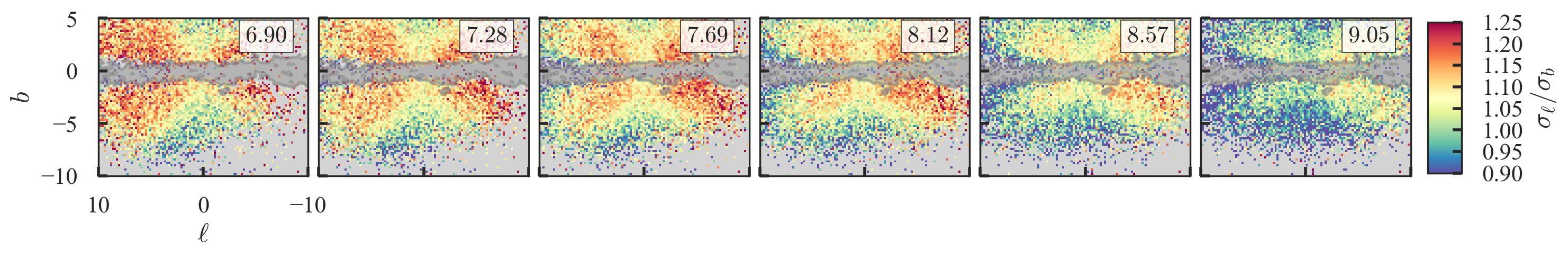}
    \caption{On-sky density and velocity dispersions of bulge giants at a series of distances. The distance in $\mathrm{kpc}$ is shown in the inset. Grey pixels correspond to uncertainties greater than: $10$ times larger than the signal in density, $7\,\mathrm{km\,s}^{-1}$ in $\sigma_i$ and $0.1$ in $\sigma_\ell/\sigma_b$. As the distance increases, the high density region moves from positive to negative Galactic longitude consistent with viewing a near end-on bar. The density peaks coincide with rising $\ell$ and $b$ dispersion as we approach the Galactic centre and then a decline on the far-side of the bar. The X-shape of the bar is visible in the dispersion ratio which is $1-1.2$ everywhere within the X.}
    \label{fig::density_slice}
\end{figure*}

We proceed to plot the density and velocity dispersions separated by distance in Fig.~\ref{fig::density_slice}.  As we sweep through the bar in distance, we observe the peak density shifting from positive to negative longitude as expected when viewing a near end-on bar. The more extended disc component is visible at positive $\ell$ in the most distant bin. Corresponding to the density, the transverse velocity dispersions are low ($70-80\,\mathrm{km\,s}^{-1}$) at the near end of the bar $\ell>0$ and then increase towards the Galactic centre (up to $\sim130\,\mathrm{km\,s}^{-1}$) before declining again on the far side of the bar ($\ell<0$, back to $70-80\,\mathrm{km\,s}^{-1}$). We also display the dispersion ratio $\sigma_\ell/\sigma_b$ in different distance slices. As we sweep through the bar, we see the $\sigma_\ell/\sigma_b\approx1.1-1.2$ X-shape appear at positive longitude and then move to negative longitudes. This region appears to be surrounded by a colder $\sigma_\ell$ envelope.

\begin{figure*}
    \centering
    \includegraphics[width=\textwidth]{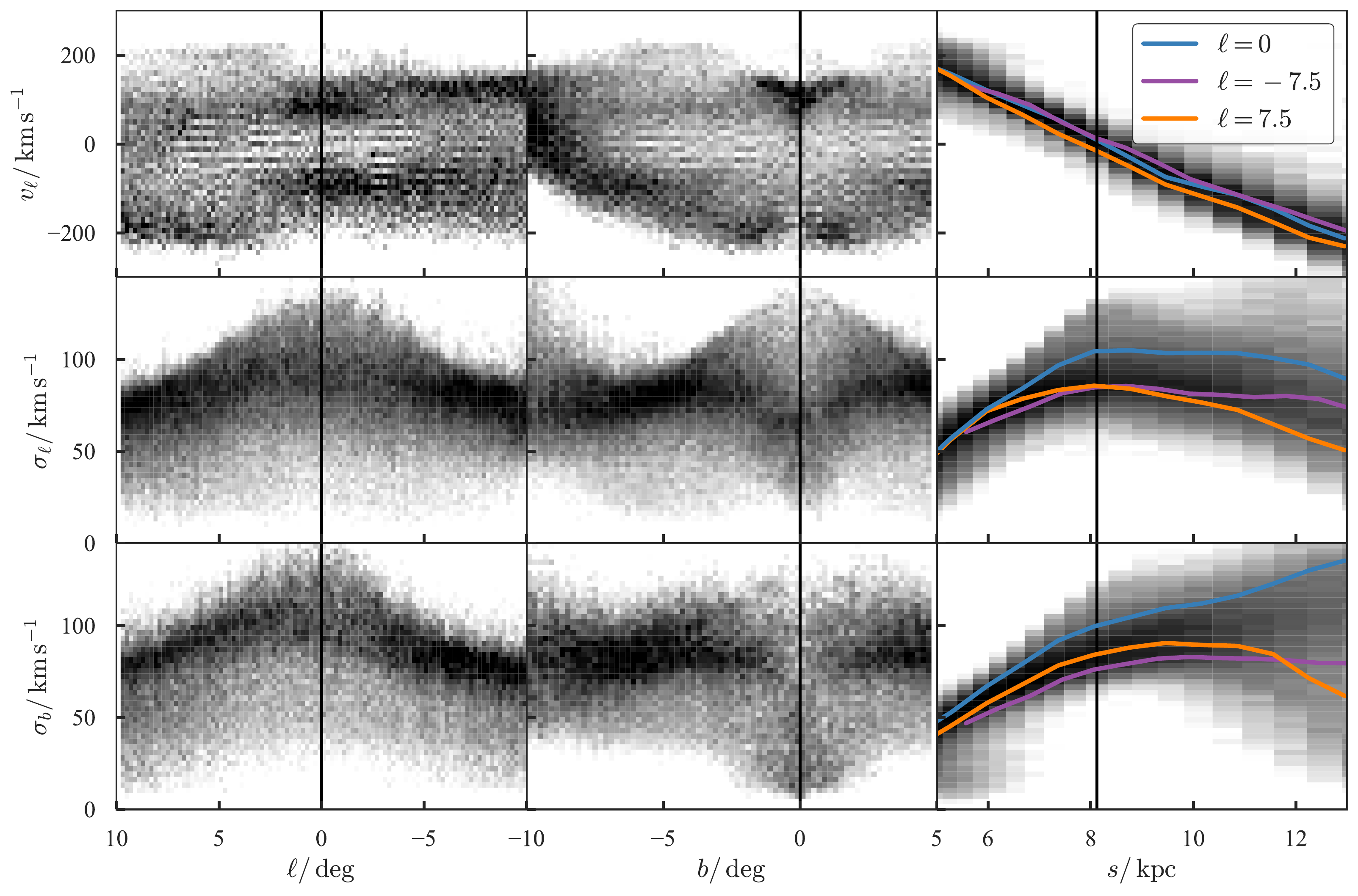}
    \caption{Column-normalized log-scaled histograms of the velocity moments against $\ell$, $b$ and distance $s$ ($v_\ell$ is corrected for the motion of the Sun). In the right column we show the running medians of $1\,\mathrm{deg}$ bins in $\ell$ centred on $0$ (blue), $-7.5$ (purple) and $7.5$ (orange). We have only used data with uncertainties in $v_\ell$ and $\sigma_i$ better than $30\,\mathrm{km\,s}^{-1}$.}
    \label{fig::hists}
\end{figure*}

For a different perspective on these results, we can bin the estimates of the moments with $\ell$, $b$ and $s$ which is displayed in Fig.~\ref{fig::hists}. Each column is normalized to one and log-scaled. We also weight by distance to produce approximately equally-spaced observations in distance instead of log-distance. In $v_\ell$, we see an asymmetric loop due the two sides of the bar. The top branch corresponds to the near-side of the bar whilst the lower branch the far-side. We see at $\ell=0$ the two branches both contribute with a gap between them due to the X-shape of the bar. The two branches are clear in the plot against $b$ and we see that at fixed $b$ we have contributions from both branches except at $b\lesssim-9\,\mathrm{deg}$, where the far-side branch doesn't reach. The striations in these plots are due Gaia systematics as they correlate with the Gaia scanning law.
As a function of distance, $v_\ell$ is nearly linear with a scatter due to different $\ell$. The near-side of the bar rotates slightly slower than the far-side in $v_\ell$ due to geometric effects. The other noticeable feature in the $v_\ell$ plots is the more isolated peaks at $|v_\ell|\approx200\,\mathrm{km\,s}^{-1}$, which is the presence of the disc population.

The two dispersions display similar trends with $(\ell,b,s)$. Both rise from approximately $70-80\,\mathrm{km\,s}^{-1}$ at $\ell=10\,\mathrm{deg}$ to maximums of $100-130\,\mathrm{km\,s}^{-1}$ at $\ell=0$ declining again towards $\ell=-10\,\mathrm{deg}$. Both dispersions are weakly asymmetric in $\ell$ due to geometric effects. In $b$, we observe similar but more symmetric behaviour. The gradients with $b$ are present but significantly flatter than those in $\ell$. Noticeably the disc population is visible in the midplane at low dispersions. Against distance, both profiles rise from $\sim40\,\mathrm{km\,s}^{-1}$ towards the Galactic centre. Beyond this, $\sigma_\ell$ instantly declines back down to near its near-field value (specifically for $\ell=7.5$)
whilst $\sigma_b$ plateaus before declining beyond $10\,\mathrm{kpc}$ (a feature also seen in our example field in Fig.~\ref{fig:example_fit}). The coloured lines show the averages over $1\,\mathrm{deg}$ bins in $\ell$. We see that the dispersion for the near-side of the bar $\ell>0$ peaks at smaller distances than the far-side $\ell<0$.

We can compare the dispersion profiles with those obtained for disc stars in e.g. \cite{Sanders2018}. The run of vertical dispersion $\sigma_b$ appears to connect onto the vertical dispersion with radius presented there although for the intermediate-age populations. This might be a reflection of selection effects of our approach (e.g. red clump stars are more likely from a younger population $\sim2\,\mathrm{Gyr}$) but perhaps more interestingly could be a reflection of the populations within the bar and when the bar buckled.

\begin{figure*}
    \centering
    \includegraphics[width=\textwidth]{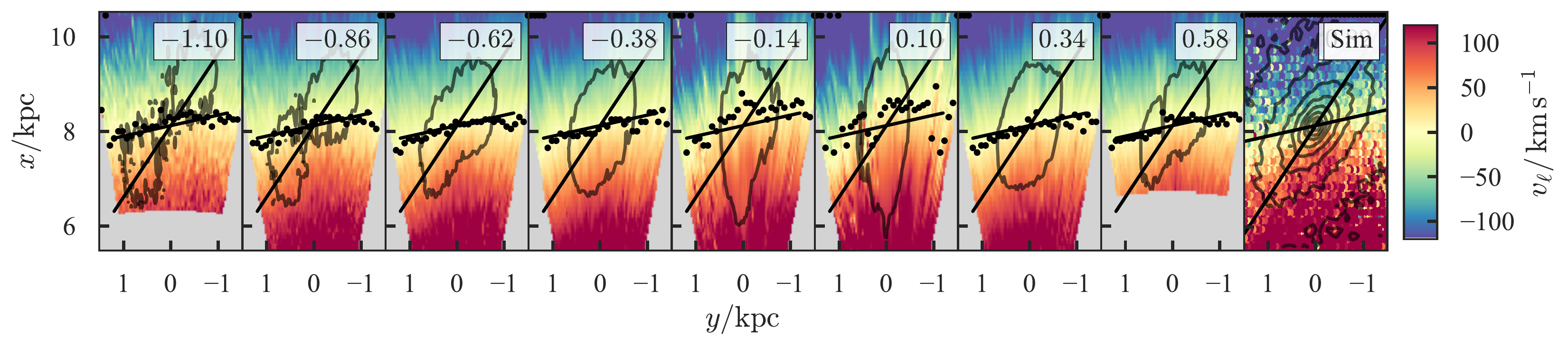}
    \caption{Top-down views of the reflex-corrected Galactic longitude velocity field in the bar for a series of Galactic height slices (labelled by the white insets). $(x,y)$ are left-handed Galactocentric Cartesian coordinates. The 80th percentile density contour is shown in black. The black dots show the location of zero velocity. The black lines are the bar major axis and a by-eye fit to the black points (angled at $12.5\,\mathrm{deg}$ to the horizontal in all panels). A rotation signal is clear but it is asymmetric in $\ell$ as expected for a non-axisymmetric structure. The broad features are reproduced by the simulation on the right, which is described in Paper II.}
    \label{fig::vl_xy}
\end{figure*}

In Fig.~\ref{fig::vl_xy}, we display the $v_\ell$ field in Galactocentric Cartesian coordinates at a range of $z$ slices. We see a clear asymmetry in the velocity field indicative of the bar (an axisymmetric rotation field in this space would be symmetric $\pm y$). We find that in all $z$ slices, the line-of-nodes (where $v_\ell=0$) is orientated at $\sim77.5\,\mathrm{deg}$ to $\ell=0$. We display a plot for results from a simulation (described in Paper II) which shows a similar relationship between the major-axis of the bar and the line-of-nodes in $v_\ell$. We also observe at high latitude (e.g. $z=860\,\mathrm{pc}$) the double peaked density field is visible and corresponds to distinctly different kinematics.

\subsection{Comparison with previous studies}
\begin{figure}
    \centering
    \includegraphics[width=\columnwidth]{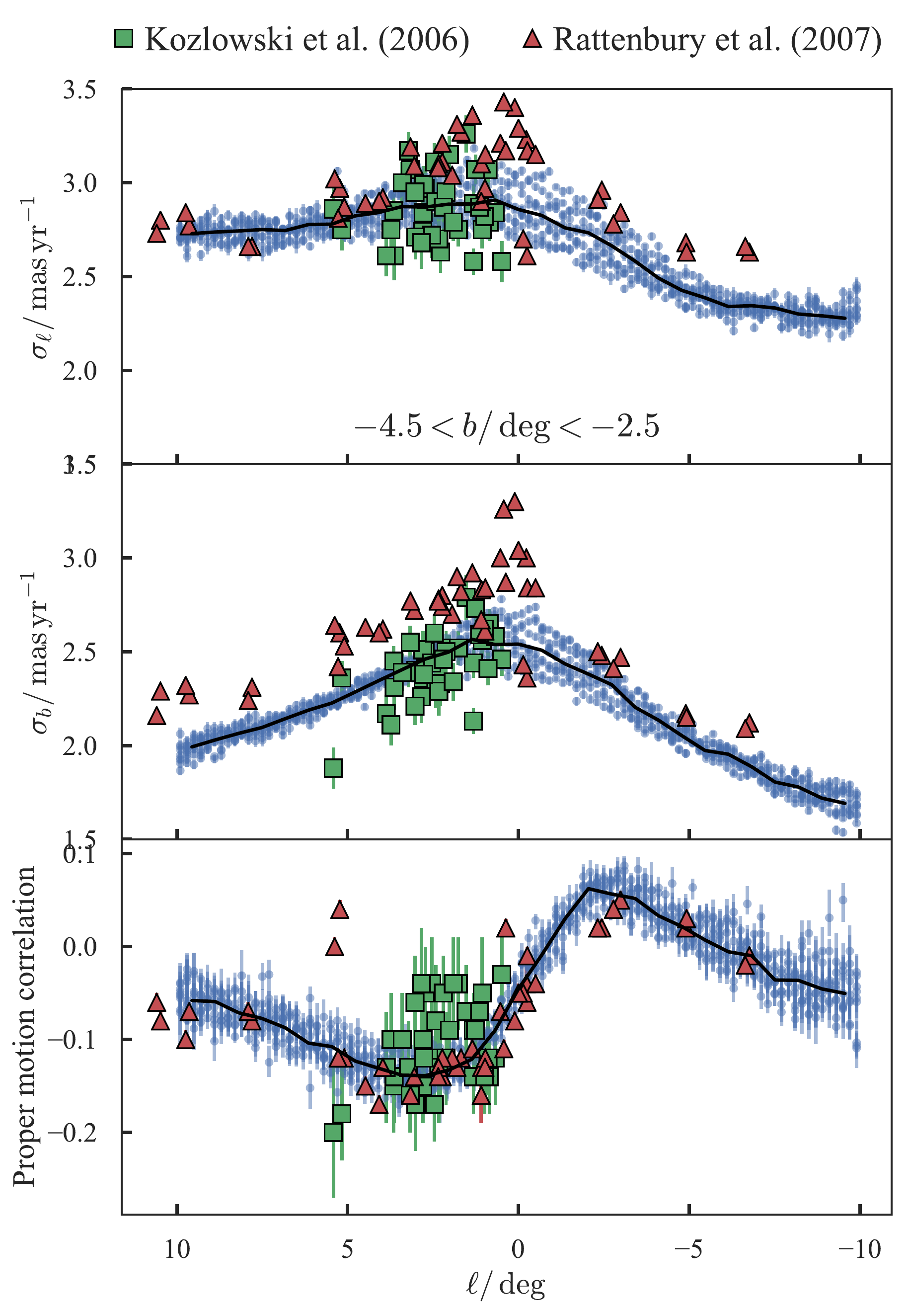}
    \caption{Comparison of this study with previous bulge proper motion studies. We compare the proper motion dispersions in $\ell$ and $b$ and the correlation between the proper motion components (blue points, black line shows median trend) with the studies of \protect\cite{Kozlowski2007} (green squares) and \protect\cite{Rattenbury2007} (red triangles). 
    }
    \label{fig:previous}
\end{figure}

Early proper motion studies of the bulge were restricted to a number of isolated low-extinction fields \citep{Spaenhauer1992,Mendez1996,Zoccali2001,Kuijken2002,Feltzing2002}. Additionally, the lack of background sources restricted these studies to relative proper motions and hence no measurement of the mean transverse velocities. The large HST programme of \cite{Kozlowski2007} extended the coverage of the bulge proper motions to 35 fields centred around $(\ell,b)=(2.5,-3)\,\mathrm{deg}$ for $16.5<I<21.5$ and \cite{Rattenbury2007} used the OGLE proper motion catalogue of \cite{Sumi2004} for $45$ fields distributed mainly along $b\approx-3.5\,\mathrm{deg}$ for $12.5<I<14.5$. Both authors measured the proper motion dispersions and the correlation between the components of the proper motion. Both studies produced consistent results finding a declining $\sigma_b$ profile with $\ell$ and variation of the proper motion correlation across the inspected fields. In Fig.~\ref{fig:previous}, we show a comparison of our proper motion measurements (not removing rotational broadening effects) within the slice $-4.5<b/\,\mathrm{deg}<-2.5$ with those of \cite{Kozlowski2007} and \cite{Rattenbury2007}. For the \cite{Rattenbury2007} measurements, the random errorbars are typically smaller than the datapoints ($\sim0.02\,\mathrm{mas\,yr}^{-1}$). Our $\sigma_\ell$ measurements agree well with those of \cite{Kozlowski2007} and are smaller than those of \cite{Rattenbury2007}. On the whole, we find very good agreement between our measurements and these previous studies. As we move away from the minor axis, our measurements decline with the fall in $\sigma_b$ steeper than that in $\sigma_\ell$. In both dispersions, there is asymmetry in $\ell$ with lower dispersions on the far-side of the bar. Our measurements are consistent with \cite{Kozlowski2007} and generally slightly smaller than \cite{Rattenbury2007}, except in $\sigma_\ell$ for $\ell>0$. This agrees with the models of \cite{Portail2017}, who found the \cite{Rattenbury2007} dispersions overpredict their model (their figure 16). This discrepancy could be causd by underestimated uncertainties in \cite{Rattenbury2007} or by the presence of contaminating populations. 

Finally, the correlation measurements agree well with both previous studies over the entire $\ell$ range, in particular with the \cite{Rattenbury2007} results. As observed previously, the correlation is smaller at negative longitude than at positive longitude. In the range $0<\ell/\,\mathrm{deg}<5$ our measured correlation is smaller (greater magnitude) than some of the \cite{Kozlowski2007} measurements.

\subsection{Comparison with spectroscopic surveys}

With proper motions, only two components of the velocity field can be mapped, unless we enforce some symmetry as in Section~\ref{Section::TriaxialStructure}. To fully map the velocity field, we require results from spectroscopic surveys. Whilst we reserve the combination of spectroscopic observations with the proper motions provided here to a separate work, we here briefly compare the transverse velocity measurements to the line-of-sight measurements across the bulge.

We consider results from five spectroscopic surveys: BRAVA, ARGOS, APOGEE, GIBS and Gaia-ESO. For BRAVA and ARGOS, we take the mean velocities and dispersions from \cite{BRAVA} and \cite{ARGOS} using our assumed solar velocities. For APOGEE, we take all fields from DR14 \citep{APOGEEDR14} within the VVV bulge footprint, remove duplicates and dwarf stars ($\log g>3.5$) and compute the mean velocity (corrected for the solar reflex) and dispersion in each field. For Gaia-ESO, we adopt a similar procedure using DR3 \citep{GaiaESO}. There is only a single field with a sufficient number of stars (at $(\ell,b)\sim(1,-4)\,\mathrm{deg}$). We remove dwarf stars if $\log g$ is available and apply a parallax cut of $\varpi<1.5\,\mathrm{mas}$ to remove nearby contaminants. For GIBS we use the radial velocity data from \cite{GIBS} and compute the reflex-corrected mean velocity and dispersion in each of the $33$ fields.

\begin{figure*}
    \includegraphics[width=\textwidth]{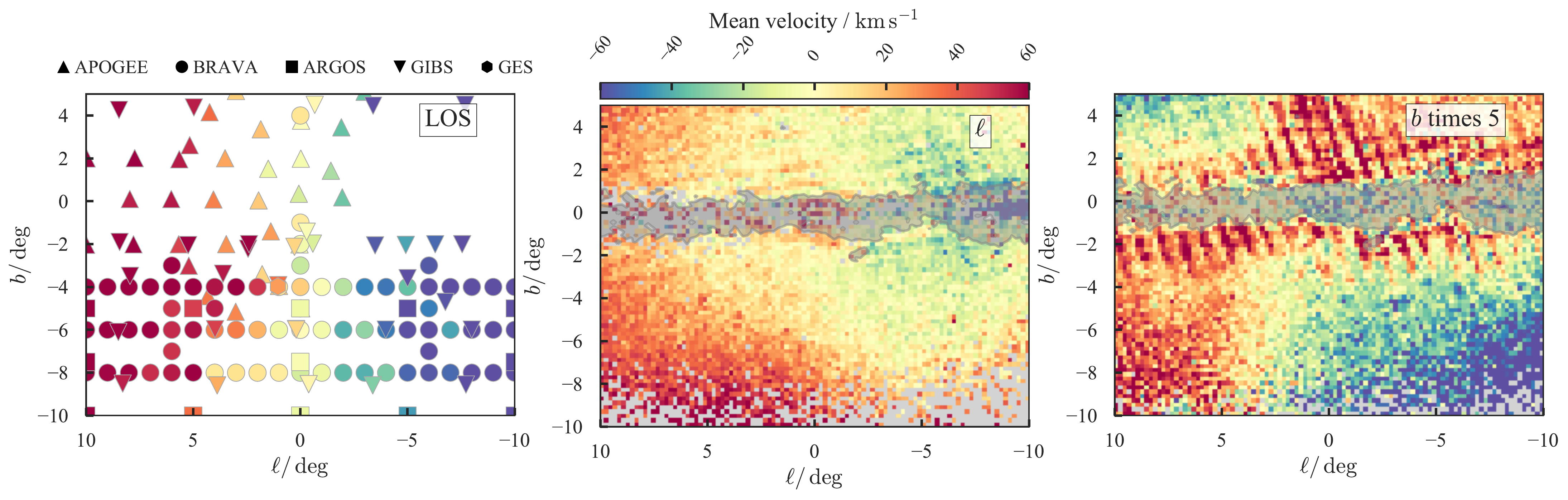}
    \includegraphics[width=\textwidth]{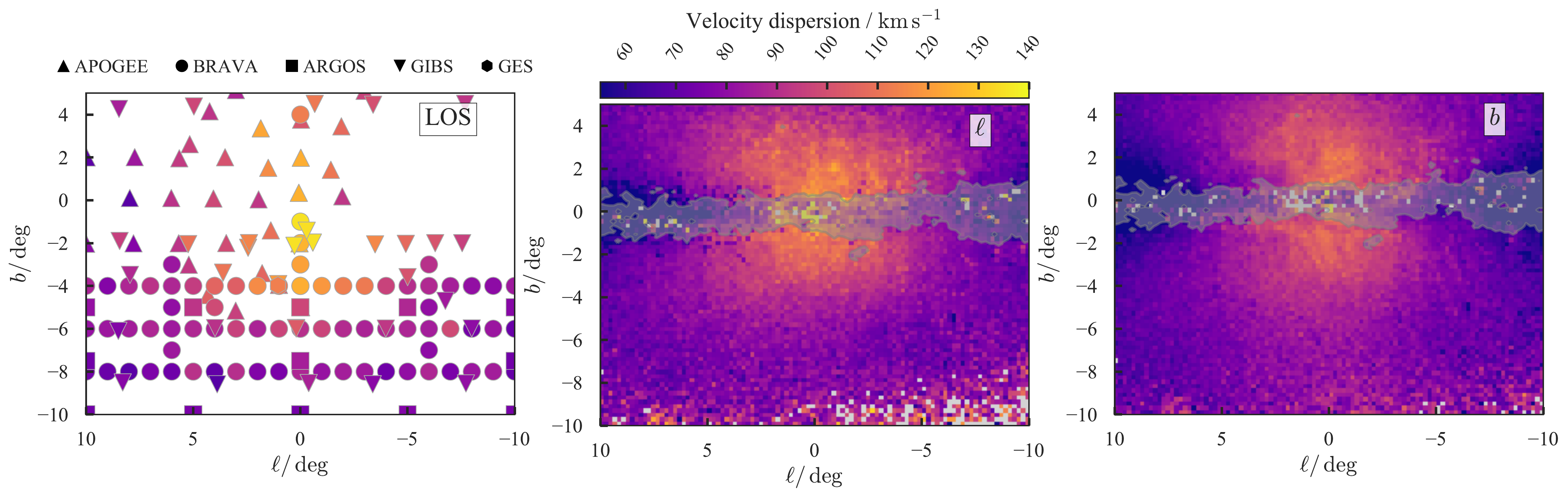}
    \caption{Mean velocity (reflex-corrected, top) and dispersion (bottom) for fields from spectroscopic surveys (left panels), and the transverse velocities from this study (middle: longitude $\ell$, right: latitude $b$). Note in the top right panel the mean velocities have been multiplied by $5$ for visibility.}
    \label{fig:spectroscopy}
\end{figure*}

In Fig.~\ref{fig:spectroscopy}, we compare the line-of-sight mean velocities and dispersions from these surveys with the results from this paper for the transverse velocity field. To attempt to compare like-to-like, we have averaged the derived transverse velocities weighted by the density profile along the line-of-sight. In the mean velocities, there is a rotation signature in all three components. The strongest signature is in the line-of-sight velocities. The longitudinal rotation signature is weaker but still visible, particularly for $\ell>0$ where the rotation is increasingly in the longitudinal direction. The rotation signature is also visible in the latitudinal direction, where as we move away from the plane there is a small rotation projection in this direction. The anticipated quadrupole signature is offset from the minor axis due to the geometry of the bar. We note that in these projections the Gaia scanning law is visible, particularly near the plane for the $b$ velocities.

In the dispersions, we see a consistent lobed structure across the three velocity components with the line-of-sight dispersion larger than the longitudinal and latitudinal. The line-of-sight and longitudinal lobes are more flattened and boxy than the slightly collimated latitudinal lobe.

\subsection{The double red clump}

\begin{figure}
    \includegraphics[width=\columnwidth]{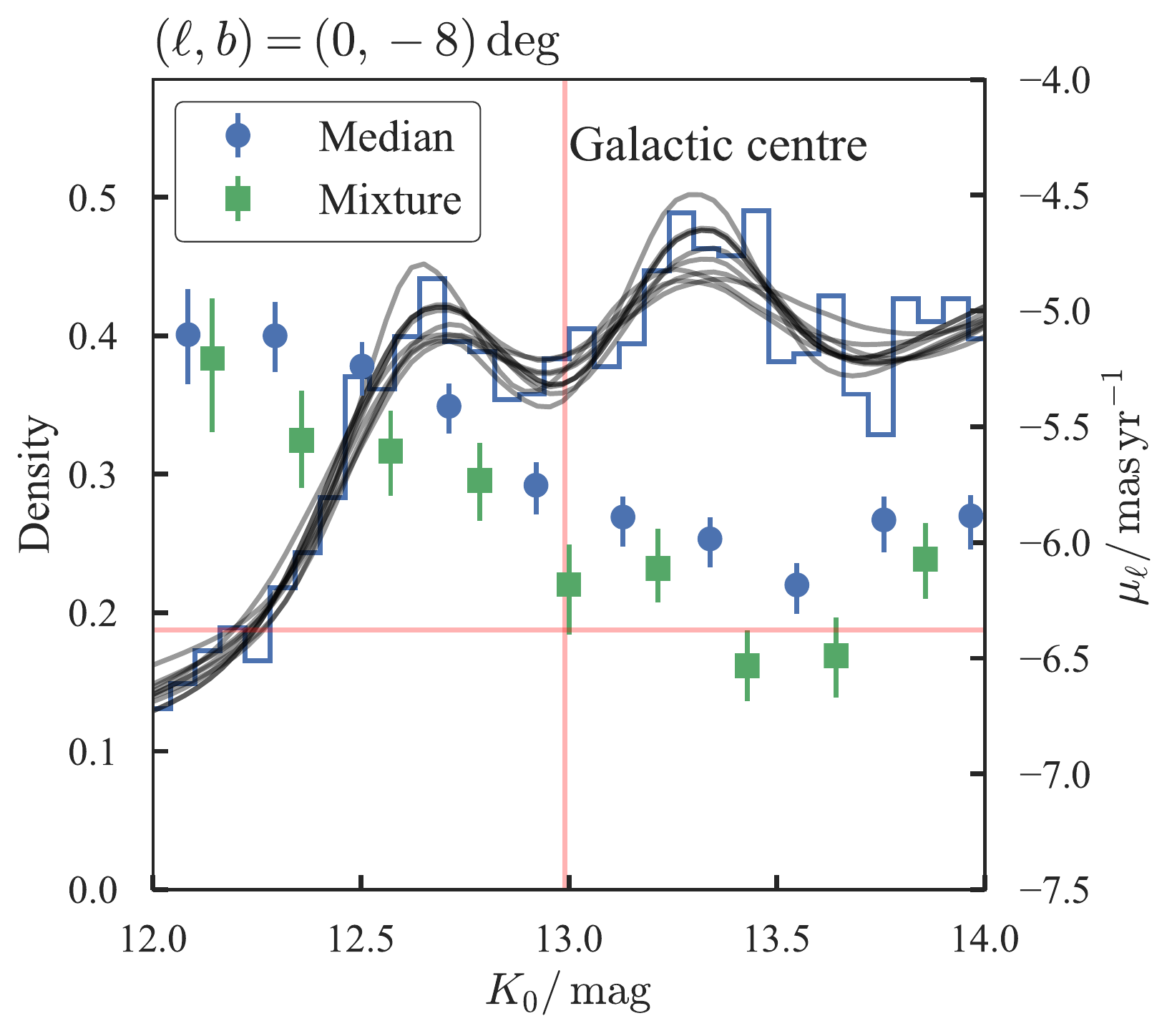}
    \caption{Kinematic difference between the split red clump for a minor axis field $(\ell,b)=(0,-8)\,\mathrm{deg}$: The number counts (blue histogram) and model (samples from which shown in black) are shown along with the average proper motion (computed using a median in blue and using a mixture model in green). Stars associated with the fainter peak have more negative $\mu_\ell$. The red lines show the magnitude of red clump stars at the Galactic centre and the proper motion of Sgr A*.}
    \label{fig:doublepeak}
\end{figure}

One of the key pieces of evidence pointing towards an X-shaped bar-bulge is the presence of a double red clump peak in the magnitude distribution of stars selected along the bulge's minor axis \citep{McWilliamZoccali}. The interpretation of this feature as two spatially-separated populations of stars has been challenged and claimed to arise from population effects \citep{Lee2015}. These claims can be refuted using the transverse velocity field for the multiple peaks. If population effects were the cause of the split, each peak would have similar kinematics. However, an X-shaped bulge scenario would give rise to differential rotation between the two peaks with the faint peak rotating in the opposite direction in the Galactocentric rest frame \citep{Poleski2013}. In Fig.~\ref{fig::vl_xy}, it is clear the transverse velocities of the two peaks present at high $|z|$ are quite distinct. However, this plot has been generated using a luminosity function to transform from $K_0$ to distance. Instead, in Fig.~\ref{fig:doublepeak} we show a field of $0.6\,\mathrm{deg}$ by $0.6\,\mathrm{deg}$ centred at $(\ell,b)=(0,-8)\,\mathrm{deg}$. We show the fitted density model which exhibits a clear double peak. We also display a running median of the $\mu_\ell$ distribution. This exhibits a clear trend with $K_0$ demonstrating the kinematic difference between the two peaks. We also fit a Gaussian mixture model where there is a fixed contaminant contribution independent of $K_0$. We generate samples from our model using \textit{emcee} \citep{emcee}. This model is shown by the green points which exhibit a slightly more significant difference in rotation velocity between the two peaks. This demonstrates conclusively that the two peaks are two spatially-separated populations. 

\section{Triaxial structure}\label{Section::TriaxialStructure}

We have inspected the bar's density and transverse velocity structure. However, the transverse velocity field is awkward to interpret as it is biased by our perspective. In this section, we shall assume triaxial symmetry for the bar-bulge allowing us to combine estimates of the velocity moments and de-project to recover the intrinsic velocity moments (and infer the missing line-of-sight velocity component). This also leads to more precise estimates of the velocity moments as we can combine up to eight different measurements.

\subsection{Triaxial density field}

We first compute the frame in which the bar appears maximally triaxial following the method of \cite{WeggGerhard2013}. For guesses of the Galactic centre distance, $R_0$, and bar angle, $\alpha$, we linearly interpolate the inferred density distributions on a rectangular grid in bar-aligned coordinates and compute the variance of the density estimates for the eight symmetry points (or fewer if the symmetry points fall outside the measured volume) divided by the mean density at the corresponding $z$. We minimise this quantity ignoring $|z|<380\,\mathrm{pc}$ to find $\alpha=23\,\mathrm{deg}$ and $R_0=8.23\,\mathrm{kpc}$ (where we note that our red clump magnitude was chosen to approximately match the \cite{GravityCollab2018} measurement of $R_0$). We find that the inference of $\alpha$ is a function of Galactic height where closer to the disc plane smaller $\alpha$ is inferred. The bar angle is slightly smaller than that of \cite{WeggGerhard2013} ($27\,\mathrm{deg}$) who modelled the overdensity relative to the smooth background. Our modelling of both disc and bulge modifies this slightly and our early models which were more similar to \cite{WeggGerhard2013} produced bar angles of $\sim28\,\mathrm{deg}$. We therefore adopt $R_0=8.12\,\mathrm{kpc}$ and $\alpha=27\,\mathrm{deg}$.

\begin{figure*}
    \centering
    \includegraphics[width=\textwidth]{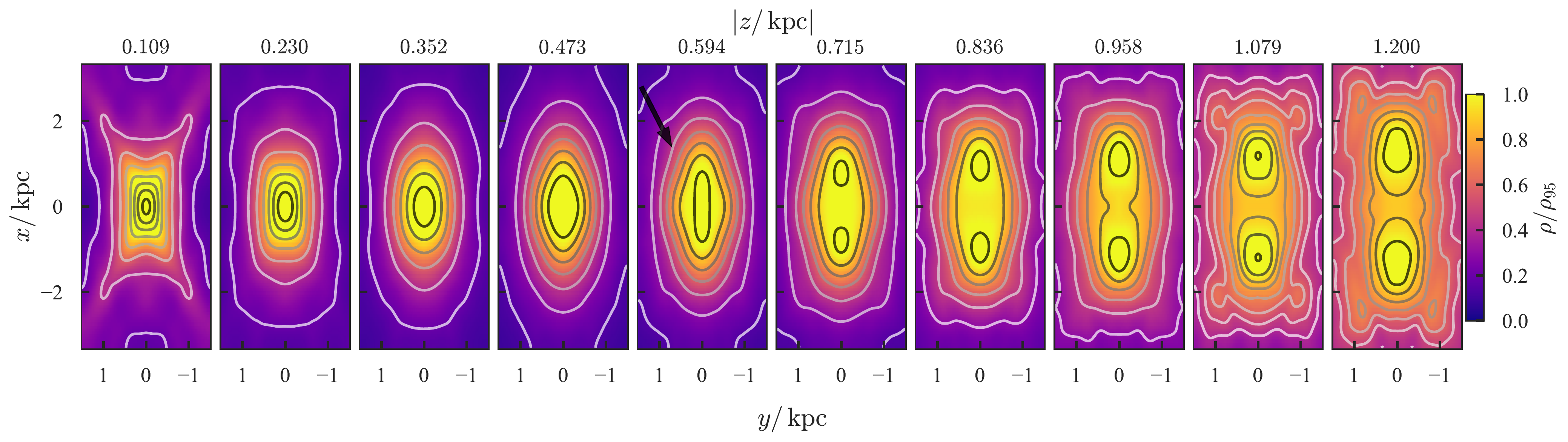}
    \includegraphics[width=\textwidth]{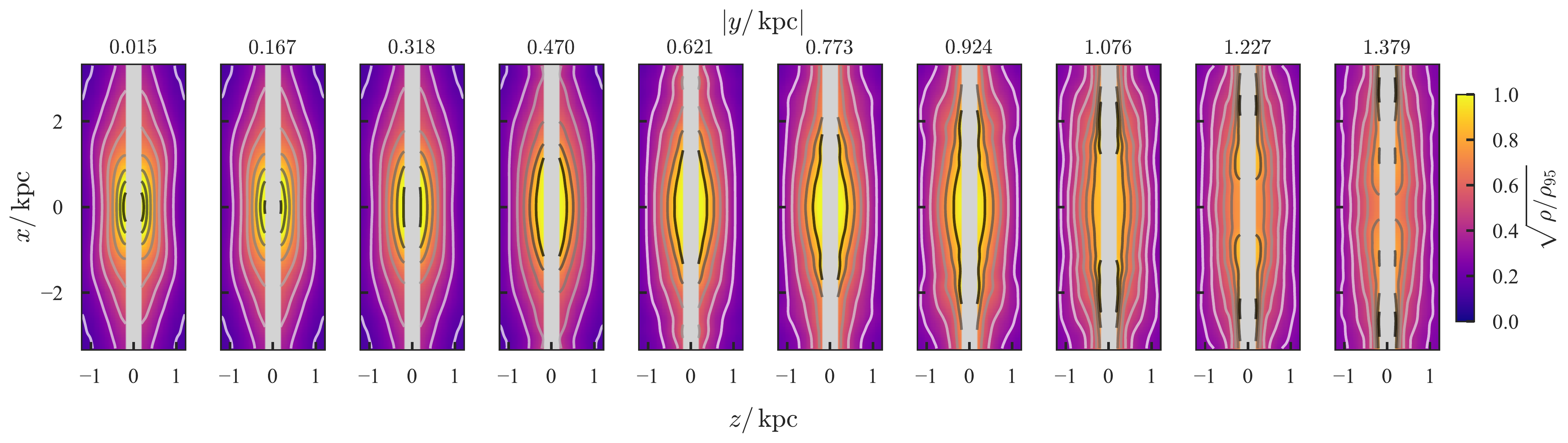}
    \includegraphics[width=\textwidth]{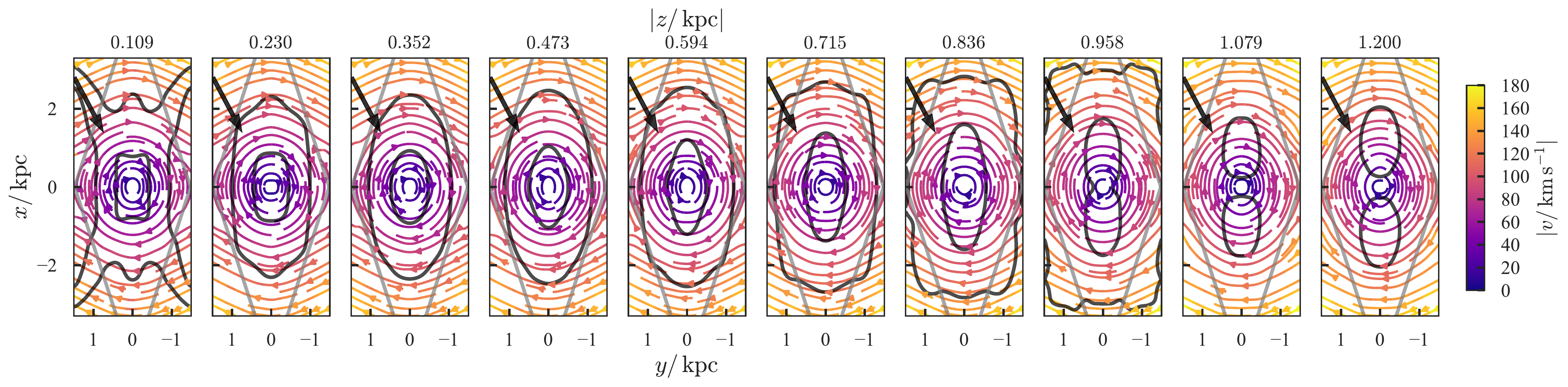}
    \includegraphics[width=\textwidth]{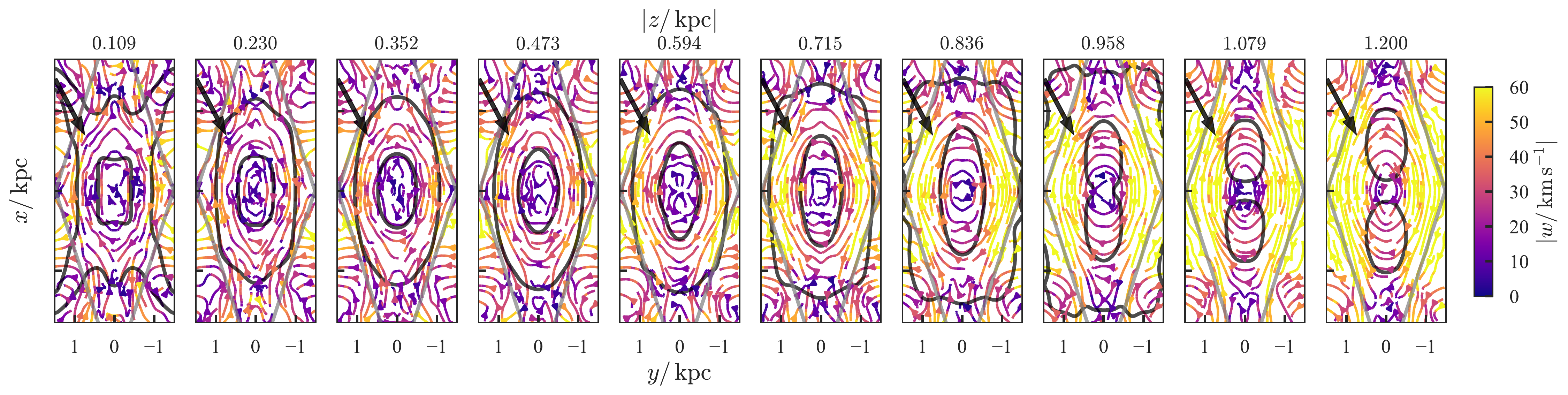}
    \caption{Triaxial density and mean velocity fields: each row of panels shows slices along the minor axis $z$ (except the second row which is sliced along the intermediate axis $y$) through the bar distribution obtained by imposing triaxiality. The top two rows show the density field, the middle two rows the mean velocity field (in the Galactocentric rest frame [top] and the frame rotating with the bar [bottom]). The black contours show two equidensity curves (at the $\sim30$th and $\sim85$th percentiles of the density) and grey lines show the region within which the recovery should be reliable.}
    \label{fig:triaxial_slices}
\end{figure*}

In this frame, we find the density field from averaging the (up to eight) equivalent points. We show slices through the resulting distribution in the top two panels of Fig.~\ref{fig:triaxial_slices}, both in the major-intermediate axis plane and major-minor axis plane. Confirming the results of \cite{WeggGerhard2013}, we see the X-shape of the bar-bulge. In the plane the bar is short and boxy with a central density peak. As we move to higher Galactic heights, the bar elongates and around $z\approx600\,\mathrm{pc}$, the density peak splits into two. This peak moves outwards as we continue to increase the Galactic height. Correspondingly, in the major-minor axis slices, the bar appears peanut-shaped towards central slices and as we move along the intermediate axis the X-shape appears. In the edge-on projections the X-shape is difficult to see at high $|y|$ due to the presence of the disc.

\subsection{Triaxial velocity field}
We can conduct a similar symmetrization procedure for the velocities but this requires more care. First, we lack one component of the velocity (the line-of-sight velocity) which we must infer from the transverse velocity measurements at a set of similar points. As we are viewing the equivalent velocities in the bar frame at slightly different angles, we can infer the missing component. Secondly, the symmetry of the velocity field of the bar is more complicated than the triaxial density structure. We assume the bar is in equilibrium and thus satisfies the continuity equation. Expressed in inertial disc coordinates aligned with the triaxial bar structure $(x,y,z)$, the continuity equation for a tracer density $\rho$ rotating with steady pattern speed $\Omega_\mathrm{p}$ reads 
\begin{equation}
    \frac{\partial\rho}{\partial t} +\nabla\cdot(\rho\bs{v}) =\Omega_\mathrm{p}\Big[y\frac{\partial\rho}{\partial x}-x\frac{\partial\rho}{\partial y}\Big]+\nabla\cdot(\rho\bs{v})=0.
\label{eqn::continuity}
\end{equation}
Here, $v_i$ are mean velocities in the inertial bar frame. Neglecting vertical flows ($v_z=0$), we write this expression as
\begin{equation}
    \frac{\partial}{\partial x}\Big(\rho(v_x+\Omega_\mathrm{p}y)\Big)=-\frac{\partial}{\partial y}\Big(\rho(v_y-\Omega_\mathrm{p}x)\Big).
\end{equation}
Under the transformation $x\rightarrow-x$, the density is triaxial so $\rho\rightarrow\rho$ and the equation reads
\begin{equation}
    -\frac{\partial}{\partial x}\Big(\rho(v_x+\Omega_\mathrm{p}y)\Big)=-\frac{\partial}{\partial y}\Big(\rho(v_y+\Omega_\mathrm{p}x)\Big).
\end{equation}
Clearly, the transformation $v_y\rightarrow-v_y$ recovers the original equation. By the same argument, there is the symmetry $y\rightarrow-y$ and $v_x\rightarrow-v_x$. This symmetry is equivalent to the $z$-component of the angular momentum being equal at all $8$ symmetry points $(\pm x,\pm y,\pm z)$. 

For the second order moments $\sigma^2_{ij}$, we introduce the potential $\Phi$ and write the Jeans equation as \citep[eq. (4.209)][]{BinneyTremaine2008}
\begin{equation}
    \rho\frac{\partial v_j}{\partial t}+\rho v_i\frac{\partial v_j}{\partial x_i}=-\rho\frac{\partial\Phi}{\partial x_j}-\frac{\partial}{\partial x_i}(\rho\sigma_{ij}^2).
\end{equation}
The mean velocity field $v_j$ is static in the rotating frame allowing us to write for $j=x$
\begin{equation}
\begin{split}
    (v_x+\Omega_\mathrm{p}y)&\frac{\partial v_x}{\partial x}+
    (v_y-\Omega_\mathrm{p}x)\frac{\partial v_x}{\partial y}+\frac{\partial\Phi}{\partial x}=\\&-\frac{1}{\rho}\Big(\frac{\partial}{\partial x}(\rho\sigma_{xx}^2)+\frac{\partial}{\partial y}(\rho\sigma_{xy}^2)+\frac{\partial}{\partial z}(\rho\sigma_{xz}^2)\Big).
\end{split}
\end{equation}
If the potential satisfies the triaxial symmetry e.g. $\Phi\rightarrow\Phi$ for $x\rightarrow-x$ (reasonable if composed of an approximately axisymmetric disc potential plus the triaxial bar potential), then under the transformation $x\rightarrow-x$, $\rho\rightarrow\rho$, $v_y\rightarrow-v_y$ and $\Phi\rightarrow\Phi$ all terms on the left-hand side change sign whilst only the first term on the right-hand side changes sign. This implies the diagonal terms of the dispersion tensor are symmetric e.g. $\sigma^2_{xx}\rightarrow\sigma^2_{xx}$ whilst the cross-terms are anti-symmetric e.g. $\sigma^2_{xy}\rightarrow-\sigma^2_{xy}$.

The matrix relating the bar-aligned coordinates (right-handed with positive $x$ corresponding to $\ell>0$ and positive $z$ towards the North Galactic Pole) to Galactic velocities $(v_\mathrm{los},v_\ell,v_b)$ is
\begin{equation}
\bs{R}
=
    \begin{pmatrix}
    -\cos\alpha&\sin\alpha&0\\
    -\sin\alpha&-\cos\alpha&0\\
    0&0&1
    \end{pmatrix}
    \begin{pmatrix}
    \cos\ell\cos b&-\sin\ell&-\cos\ell\sin b\\
    \sin\ell\cos b&\cos\ell&-\sin\ell\sin b\\
    \sin b&0&\cos b
    \end{pmatrix}.
\end{equation}
We seek the velocities in the positive octant e.g. $\bs{v}(x>0,y>0,z>0)=(v_x,v_y,v_z)$ using (reflex-corrected) data $v'_\ell,v'_b$ (and uncertainties $\Delta_\ell,\Delta_b$) from all octants. Additionally, we recover the `true' Galactic velocities $\bs{v}_g=(v_\mathrm{los},v_\ell,v_b)$. We use the sequential quadratic programming algorithm (SLSQP) implemented in \texttt{scipy.optimize} to minimise 
\begin{equation}
    \sum_{i=1}^8\frac{(v_{\ell,i}-v'_{\ell,i})^2}{\Delta_{\ell,i}^2}+\frac{(v_{b,i}-v'_{b,i})^2}{\Delta_{b,i}^2}
\end{equation}
subject to the constraints
\begin{equation}
    \mathrm{sgn}_i\cdot\bs{v}-\bs{R}(\ell_i,b_i)\cdot\bs{v}_{gi}=\bs{0}\mathrm{\,for\,all\,octants\,}i.
\end{equation}
Here, $i$ indexes the octants (where we only consider octants with data) and $\mathrm{sgn}_i$ is a 3-vector of $\pm1$ enforcing the previously derived symmetry (for $z$ we use the symmetry $z\rightarrow-z$ and $v_z\rightarrow-v_z$ although we expect $v_z=0$).

We symmetrise the dispersion field in a similar way. Using unprimed to denote `true', primed observed and $\Delta$ the uncertainties, we minimise
\begin{equation}
    \sum_{i=1}^8\frac{(\sigma^2_{\ell,i}-\sigma^{2'}_{\ell,i})^2}{\Delta_{\sigma\ell,i}^2}+\frac{(\sigma^2_{b,i}-\sigma^{2'}_{b,i})^2}{\Delta_{\sigma b,i}^2}+\frac{(\rho^{}_{\ell b,i}-\rho'_{\ell b,i})^2}{\Delta_{\rho,i}^2},
\end{equation}
subject to the constraints
\begin{equation}
    \mathrm{sgnS}_{i}\cdot\bs{\sigma}^2-\bs{R}(\ell_i,b_i)\cdot\bs{\sigma}^2_{gi}\cdot \bs{R}^T(\ell_i,b_i)=\bs{0}\mathrm{\,for\,all\,octants\,}i,
\end{equation}
where $\bs{\sigma}^2_{gi}$ is the true dispersion tensor in Galactic coordinates, $\bs{\sigma}^2$ the tensor in Cartesian bar-aligned coordinates and $\mathrm{sgnS}_{i}$ a tensor of $\pm1$ enforcing the required symmetry. Again we use SLSQP to find the optimum. However, in this case we find there are local minima and the result depends sensitively on the initial guess of $\sigma^2_\mathrm{los}$. We therefore run SLSQP on a small grid of initial guesses for $\sigma^2_\mathrm{los}$. 

We carry out these procedures on a rectangular grid in bar-aligned coordinates interpolating the required quantities at each position. We take $\rho_{\ell b}$ as constant in distance for each on-sky position (due to the limitations of our modelling procedure in Section~\ref{Section::Method}).

\begin{figure*}
    \centering
    \includegraphics[width=\textwidth]{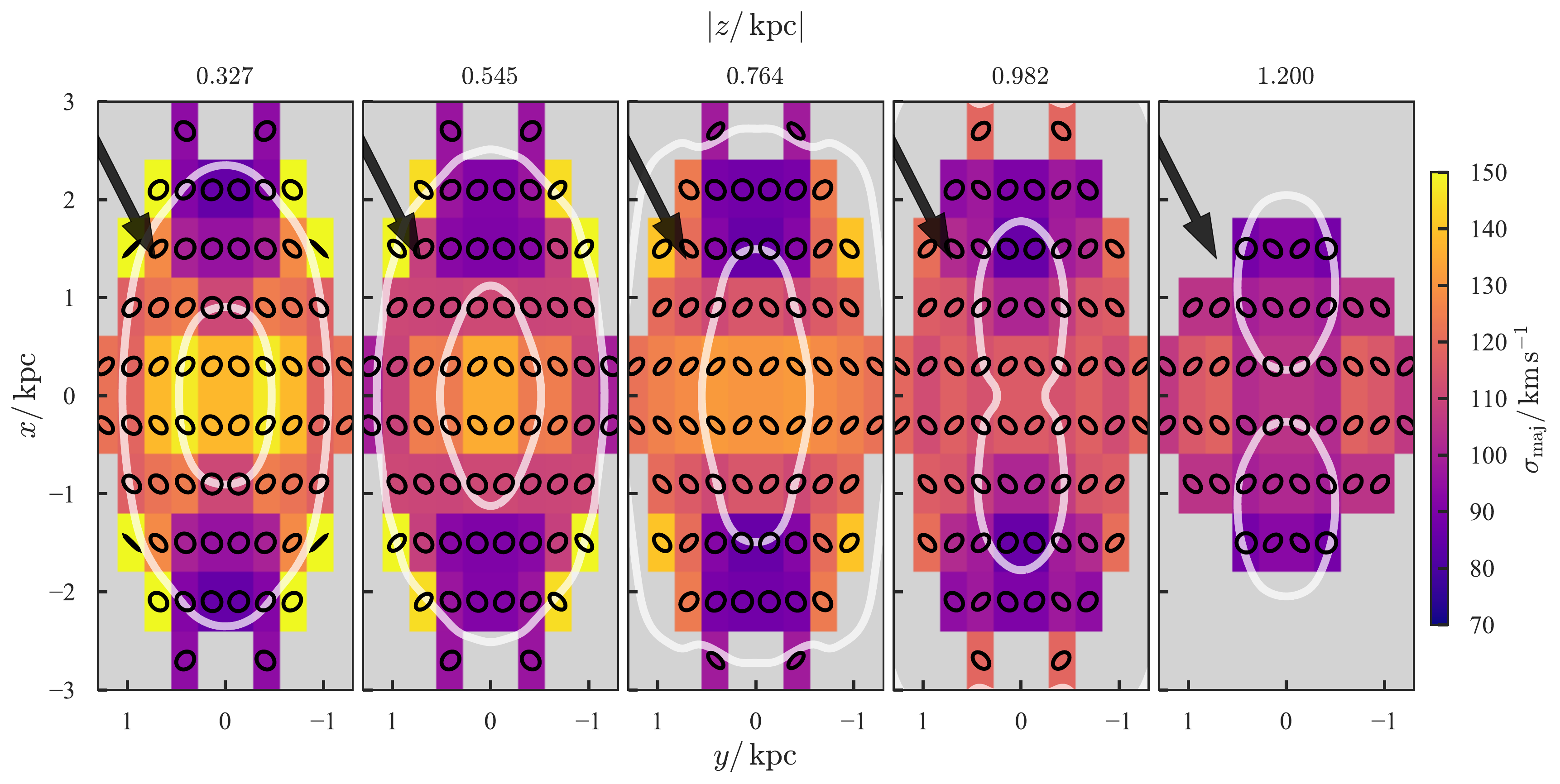}
    \includegraphics[width=\textwidth]{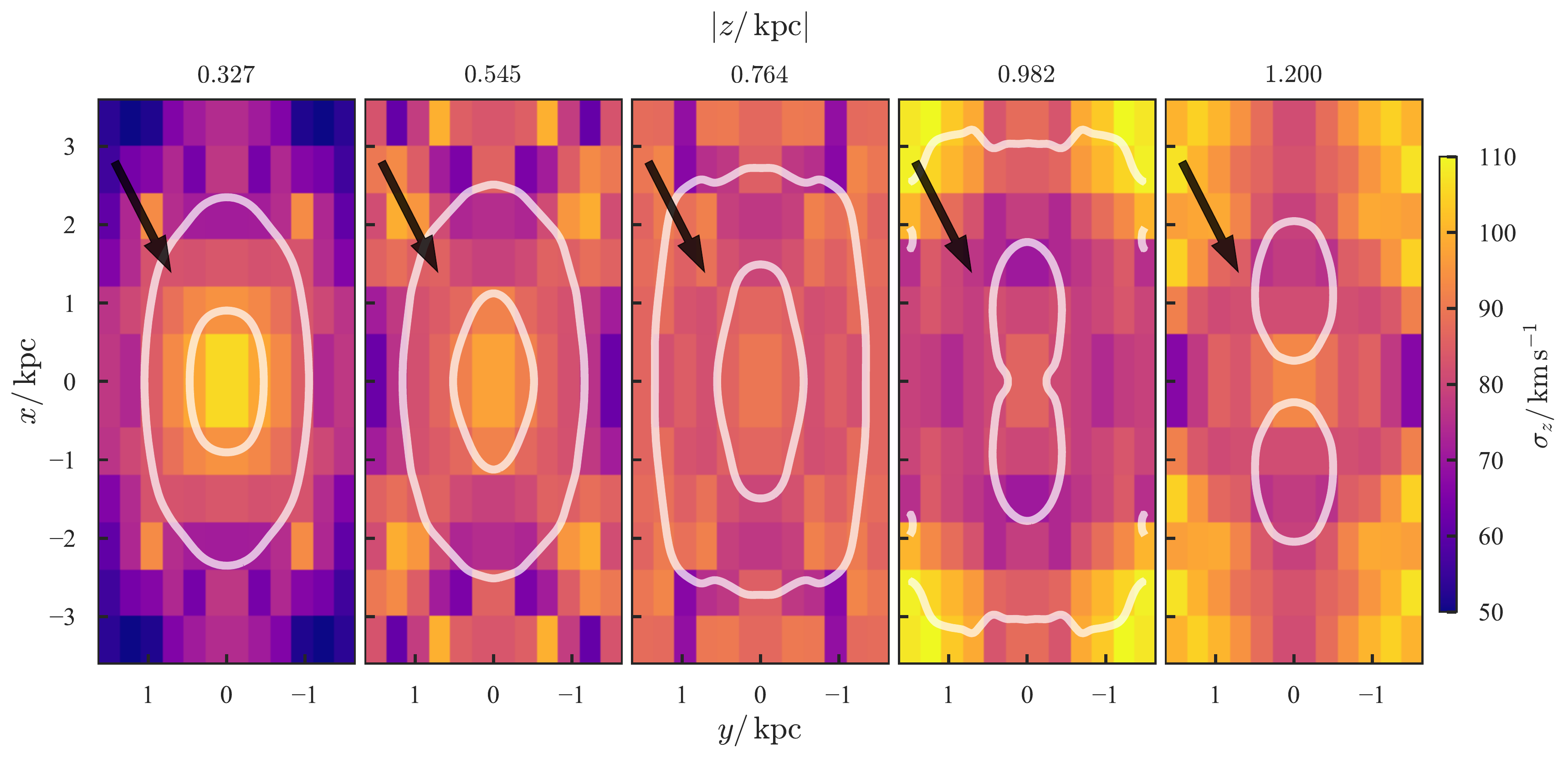}
    \caption{Triaxial dispersion fields: each panel is a slice in Galactic height $z$ (labelled above plot) through the dispersion field in the bar frame. The top row of panels is coloured by the major axis length of the in-plane dispersion tensor (only showing pixels where more than two symmetry points are observed in the $z$ slice) and the shapes of the velocity ellipsoids at each location are overlaid. The bottom row is coloured by the vertical dispersion. Overlaid in white are equi-density contours. The black arrow shows the viewing direction.
    }
    \label{fig:triaxial_dispersion_slices}
\end{figure*}

\subsubsection{Tests of method}

In Appendix~\ref{Appendix} we test our method on a mock triaxial velocity field. We find that the mean velocities are well recovered and when three or four equivalent points in a single $z$ slice are observed the mean velocity uncertainties are similar to the input uncertainties. Outside this region, the uncertainties are $\sim3$ times larger but not significantly biased in the average. We obtain a similar result for the dispersion field but when we have fewer than three equivalent points in a $z$ slice the $x$ and $y$ dispersions can be signifcantly biased.

As a further test of our method we can compare the recovery of the line-of-sight mean velocity and dispersion to the spectroscopic survey data. For each spectroscopic field, we interpolate the line-of-sight velocity moments at a set of distances and find the mean moments weighted by $s^2\rho$ (for the line-of-sight dispersion we sum the mean dispersion with the dispersion in the mean). For the dispersion, we only use points reconstructed from three or more symmetry points (see Appendix~\ref{Appendix}). We show the results in Fig.~\ref{fig:spectro_recovery}. The agreement of the mean velocities is satisfying, particularly as we have not attempted to match the distance distribution of the spectroscopic studies. The velocity dispersion match is not as good. At the high dispersion end, the match is adequate although with large scatter, but at the low dispersion end the reconstruction dispersion us approximately $15\,\mathrm{km\,s}^{-1}$ higher than the spectroscopic dispersion. This is possibly a shortcoming of only using fields with more than three symmetry points used in their reconstruction. This limits us to more central fields and so overestimates the dispersion which declines with distance from the Galactic centre. 

\begin{figure}
    \centering
    \includegraphics[width=\columnwidth]{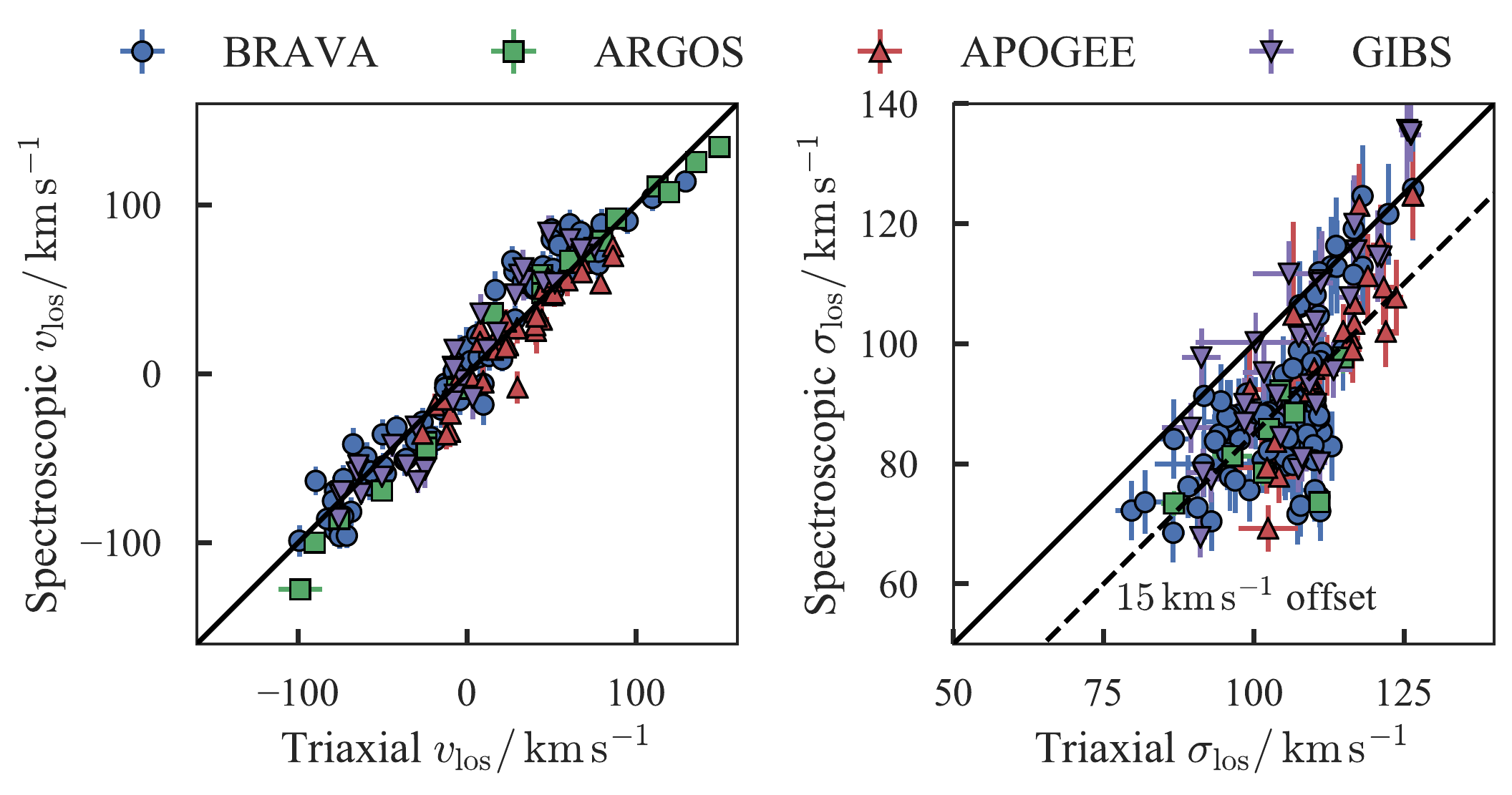}
    \caption{Comparison of the measured spectroscopic velocity moments with our triaxial velocity field measured from solely proper motions. Solid black is a one-to-one line and dashed black is offset such that the recovery is $15\,\mathrm{km\,s}^{-1}$ higher than the spectroscopic result.}
    \label{fig:spectro_recovery}
\end{figure}

\subsubsection{Results}

In the bottom panels of Fig.~\ref{fig:triaxial_slices}, we show the velocity field in the bar frame, both inertial and co-rotating. We observe a clear cylindrical rotation signature in the inertial velocities with the magnitude of the velocities increasing smoothly with radius from the centre. As we move up through the bar, the rotation decreases in amplitude. When we move to the frame co-rotating with the bar \citep[using a pattern speed of $40\,\mathrm{km\,s}^{-1}\mathrm{kpc}^{-1}$,][]{Portail2017},
we see the rotation is not purely cylindrical but there is net flow along the bar. The shape of the velocity field is pinched approximately tracing the density field. We interpret this as the effect of the x1 orbits -- the dominant bar-supporting orbits which rotate in the prograde sense \citep{BinneyTremaine2008}. 
At higher Galactic heights the amplitude of the streaming along the bar increases.

In Fig.~\ref{fig:triaxial_dispersion_slices}, we display the in-plane and vertical velocity dispersion field recovered with our method. The results appear reasonable everywhere within the region expected by the analysis presented in Appendix~\ref{Appendix}. We observe the in-plane major axis dispersion decays with Galactic height and along the bar major axis, whilst along the bar intermediate axis the decay is a lot weaker or not clear. The structure of the field appears to trace the density profile and there is a suggestion of colder in-plane dispersion at the tips of the X-shaped bar (in the $0.982$ panel). 
The velocity ellipsoid appears to be preferentially tangentially aligned nearly everywhere. At higher latitudes, the axis ratio is smaller than at lower latitudes. At higher latitudes the field becomes less structured.

The vertical dispersion exhibits similar features to the in-plane dispersion decaying both vertically and radially. At low latitudes the contours of equal velocity dispersion have an elliptical shape. At higher latitudes, the dispersion is more uniform in $x$ and $y$. At large radii at high latitude, the dispersion is large. This is possibly due to the disc component or due to unreliable dispersion recovery as these field occur in the region where we have only two symmetry points. However, our tests show the vertical dispersion is well recovered even in these regions as this information is almost solely in $\sigma_b$.

\section{Conclusions}\label{Section::Conclusions}

We have extracted the density and kinematics of $\sim45$ million bulge giants from the near-infrared VVV survey complemented with proper motions from Gaia DR2 and the VIRAC catalogue. We have probabilistically measured the transverse velocity field as a function of distance from the proper motion distributions as a function of magnitude. We have used the transverse velocity field to construct the full 3d kinematics under the assumption of triaxiality finding good agreement with the corresponding spectroscopic observations. Our conclusions are as follows:
\begin{itemize}
    \item The transverse velocity distribution is consistent with viewing a near end-on bar with the near end at positive longitudes. The proper motion dispersions are in general larger at positive longitude. When modelling the dispersions with distance, we find both dispersions rise from $\sim50\,\mathrm{km\,s}^{-1}$ at positive latitudes at distances of $\sim6\,\mathrm{kpc}$ to central dispersions of $\sim130\,\mathrm{km\,s}^{-1}$ and then declines back down to $\sim50\,\mathrm{km\,s}^{-1}$ at negative latitudes at distances of $\sim11\,\mathrm{kpc}$. 
    \item The on-sky dispersions decline with $|\ell|$ and $|b|$. They produce and extend trends seen in previous studies. The $\ell$ dispersion forms a more boxy profile than the more collimated $b$ dispersion. The decline along the minor axis axis flattens beyond $5\,\mathrm{deg}$.
    \item There is a large-scale X-structure in the on-sky $\sigma_\ell/\sigma_b$ maps with typical values on the minor axis of $1.1$ increasing to $1.2-1.3$ outside $|\ell|\approx3\,\mathrm{deg}$ and then increasing significantly in the disc plane. Removing the rotational broadening, we find the dispersion ratio decreases from $1.1$ $\sim0.9$ along the minor axis away from the Galactic centre, but the large-scale X morphology persists. Slicing through in distance shows this X is orientated along the bar.
    \item The $\ell,b$ proper motion correlation has a clear on-sky quadrupole signature with amplitude $\sim0.2$ and is approximately radially-aligned across the bulge region. The correlation is weaker at negative $\ell$ due to geometric effects.
    \item The cylindrical rotation signature observed in the spectroscopic surveys of the bulge is confirmed by the transverse velocity field. The $\ell$ transverse velocity field is clearly asymmetric in $\ell$ and corresponds well to a dynamically-formed bar model. The line-of-nodes is orientated at approximately $77.5\,\mathrm{deg}$ to the $\ell=0$ line.
    \item The transverse velocity dispersions exhibit a similar lobed structure to that seen in the spectroscopic surveys of the bulge. The amplitude of the line-of-sight dispersion is typically larger than the longitudinal dispersion which in turn is larger than the latitudinal dispersion. 
    \item The double peak magnitude distribution of minor axis bulge fields displays different kinematics for the brighter and fainter peaks. This is confirmation of the X-shaped bulge where the brighter peak is rotating differentially with respect to the fainter peak.
    \item The 3d rotation field constructed by assuming triaxiality exhibits a near-cylindrical structure but in the co-rotating frame streaming along the bar is evident and indicative of x1 orbits. The corresponding in-plane velocity dispersion field exhibits tangential bias across most of the bulge region.
\end{itemize}

The results presented in this paper provide constraints for Milky Way bar models. In a companion paper (Sanders, Smith \& Evans, submitted, Paper II), we use the results obtained in this work to estimate the pattern speed of the bar using the continuity equation. Detailed kinematics are an essential part of dynamical modelling, and hence extraction of the underlying Galactic potential in the central regions of the Galaxy. For instance, the provided transverse velocity moments can be used to constrain Made-to-Measure models of the Galactic bar-bulge \citep[e.g.][]{Portail2017}. To-date proper motions have primarily been used as an a posteriori check of any modelling that has been fitted using spectroscopic observations. Proper motions open the possibility of stricter kinematic constraints and hence tighter estimates of the bulk properties of the bar. However, more data is expected to test the assumptions required for dynamical modelling. For instance, the bar is unlikely to be truly triaxially symmetric. Full 3D kinematics from a combination of proper motions and spectroscopic data will allow stricter tests of the equilibrium nature of the bar. Furthermore, the combination of proper motions with spectroscopic surveys allows for the separation of different bar populations by chemical abundances \citep{Portail2017} providing constraints on the formation of the bar.

\section*{Acknowledgements}
JLS thanks the Science and Technology Facilities Council, the Leverhulme Trust, the Newton Trust and Christ's College, Cambridge for financial support. 
We acknowledge the simultaneous work by Clarke et al. (2019), who also used an absolute proper motion catalogue derived from VVV and Gaia DR2 to study the kinematics of the bulge. The authors of both publications were aware of each others work, but arrived at their conclusions independently.
We acknowledge useful conversations with Chris Wegg, Jonathan Clarke, Ortwin Gerhard, Eugene Vasiliev and the Cambridge Streams Group.

Based on data products from observations made with ESO Telescopes at the La Silla or Paranal Observatories under ESO programme ID 179.B-2002. This work has made use of data from the European Space Agency (ESA) mission
{\it Gaia} (\url{https://www.cosmos.esa.int/gaia}), processed by the {\it Gaia}
Data Processing and Analysis Consortium (DPAC,
\url{https://www.cosmos.esa.int/web/gaia/dpac/consortium}). Funding for the DPAC
has been provided by national institutions, in particular the institutions
participating in the {\it Gaia} Multilateral Agreement. This publication makes use of data products from the Two Micron All Sky Survey, which is a joint project of the University of Massachusetts and the Infrared Processing and Analysis Center/California Institute of Technology, funded by the National Aeronautics and Space Administration and the National Science Foundation. Funding for the Sloan Digital Sky Survey IV has been provided by the Alfred P. Sloan Foundation, the U.S. Department of Energy Office of Science, and the Participating Institutions. SDSS-IV acknowledges support and resources from the Center for High-Performance Computing at the University of Utah. The SDSS web site is www.sdss.org. SDSS-IV is managed by the Astrophysical Research Consortium for the  Participating Institutions of the SDSS Collaboration including the 
Brazilian Participation Group, the Carnegie Institution for Science, 
Carnegie Mellon University, the Chilean Participation Group, the French Participation Group, Harvard-Smithsonian Center for Astrophysics, 
Instituto de Astrof\'isica de Canarias, The Johns Hopkins University, 
Kavli Institute for the Physics and Mathematics of the Universe (IPMU) / 
University of Tokyo, the Korean Participation Group, Lawrence Berkeley National Laboratory, 
Leibniz Institut f\"ur Astrophysik Potsdam (AIP),  
Max-Planck-Institut f\"ur Astronomie (MPIA Heidelberg), 
Max-Planck-Institut f\"ur Astrophysik (MPA Garching), 
Max-Planck-Institut f\"ur Extraterrestrische Physik (MPE), 
National Astronomical Observatories of China, New Mexico State University, 
New York University, University of Notre Dame, 
Observat\'ario Nacional / MCTI, The Ohio State University, 
Pennsylvania State University, Shanghai Astronomical Observatory, 
United Kingdom Participation Group,
Universidad Nacional Aut\'onoma de M\'exico, University of Arizona, 
University of Colorado Boulder, University of Oxford, University of Portsmouth, 
University of Utah, University of Virginia, University of Washington, University of Wisconsin, 
Vanderbilt University, and Yale University.




\bibliographystyle{mnras}
\bibliography{bibliography} 



\appendix
\section{Reconstructing a mock triaxial velocity field}\label{Appendix}
In this appendix, we test our method for recovering a triaxial velocity field from transverse velocity data. We construct a very simple mean velocity and dispersion field according to
\begin{equation}
\begin{split}
    v_x(\bs{x}) &= v_0\frac{a^2y}{\sqrt{b_r^4x^2+a^4y^2}},\>\>
    v_y(\bs{x}) = v_0\frac{-b_r^2x}{\sqrt{b_r^4x^2+a^4y^2}},\\
    \sigma_i(\bs{x}) &= \sigma_{i0}\exp\Big(-\frac{x^2}{a^2}
    -\frac{y^2}{b^2}
    -\frac{z^2}{c^2}\Big),
\end{split}
\end{equation}
where $(a,b,c)=4(1,0.7,0.6)\,\mathrm{kpc}$, $b_r=2.1\,\mathrm{kpc}$,  $\bs{\sigma}_{0}=(140,100,100)\,\mathrm{km\,s}^{-1}$, $v_0=150\,\mathrm{km\,s}^{-1}$. $v_z$ and the cross-terms in the dispersion tensor are set to zero. In this model the dispersions and the total velocities are constant on ellipses for fixed $z$. This by-eye produces mocks that resemble the data. We sample the model on a $20$ by $20$ by $20$ grid in $-3.5<x/\,\mathrm{kpc}<3.5$, $-1.5<y/\,\mathrm{kpc}<1.5$ and $-1<z/\,\mathrm{kpc}<1$ applying the on-sky VVV bulge footprint. We transform to observables $\langle\mu_\ell\rangle$ etc. using an assumed bar angle of $28\,\mathrm{deg}$ and we add uncertainties in the mean velocities and dispersions of $20\,\mathrm{km\,s}^{-1}$ and  $10\,\mathrm{km\,s}^{-1}$ respectively and in the correlation of $0.03$ (typical values from our fits to the data) . We then apply the methodology outlined in Section~\ref{Section::TriaxialStructure}.

\begin{figure}
    \includegraphics[width=\columnwidth]{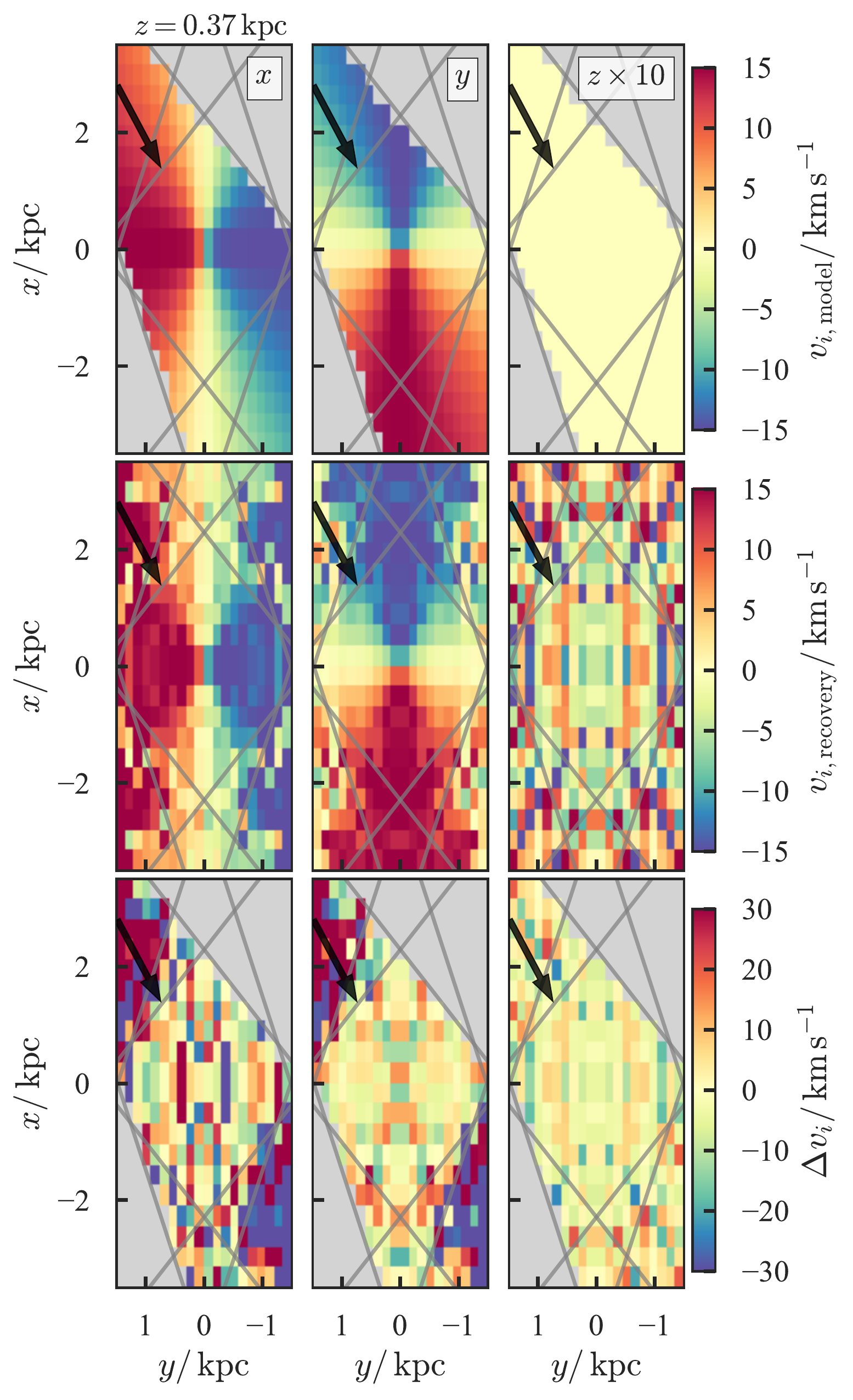}
    \caption{Recovery of the mean velocity field from mock transverse velocity observations for a slice at $z=0.37\,\mathrm{kpc}$. Each column corresponds to a different velocity component in the bar frame ($x$, $y$, $z$). The top row is the mock `truth', the middle the recovery and the bottom the residuals. In the top two rows we have multiplied the $z$ component by $10$. The grey lines show the selection volume limits reflected about the symmetry axes and  the arrows show the observation direction.}
    \label{fig:mean_field}
\end{figure}

\begin{figure}
    \centering
    \includegraphics[width=.48\columnwidth]{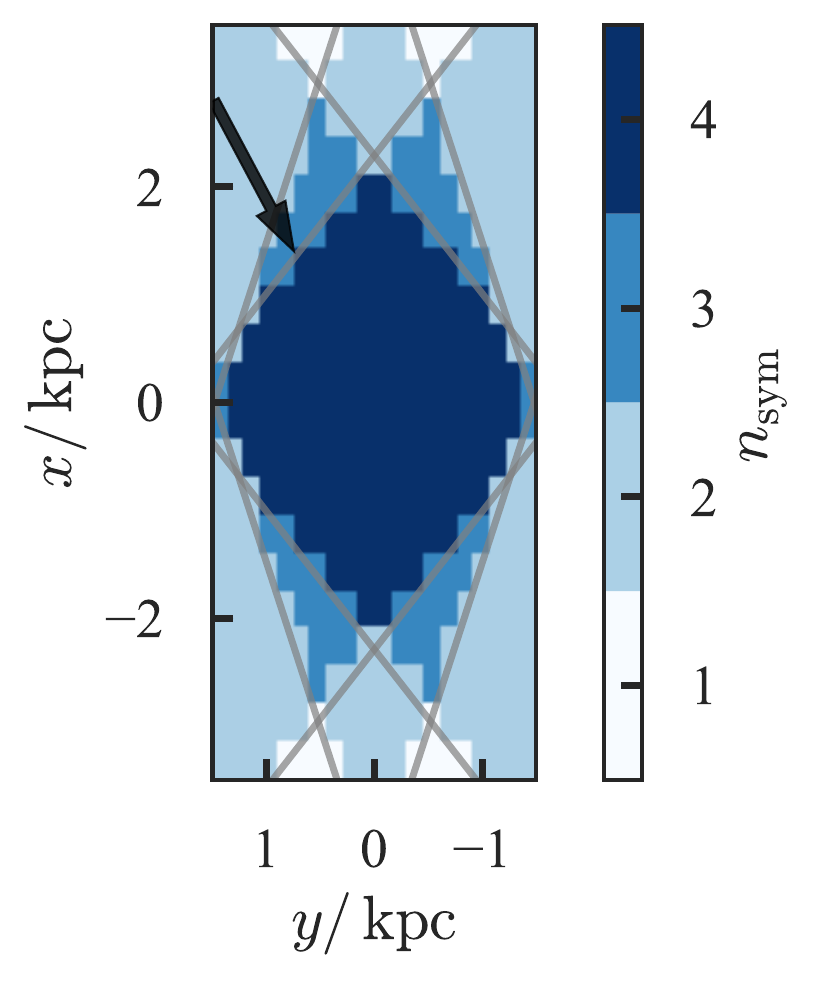}
    \caption{Number of equivalent points in the bar frame observed by the VVV footprint in a slice in $z$. The grey lines delineate the different regions. The arrow shows the observation direction.}
    \label{fig:sym_points}
\end{figure}

A slice through the results of the mean velocity field fits are shown in Fig.~\ref{fig:mean_field}. In the top row we display the true input field, middle the recovery and bottom the difference (recovery minus truth). We have overlaid grey lines showing the selection volume reflected in the symmetry axes. We observe that the overall structure is well reproduced with smaller uncertainties at the centre than in the outskirts and nowhere are the results significantly biased. We can understand this structure by inspecting the number of equivalent points used at each $(x,y)$ which is displayed in Fig.~\ref{fig:sym_points}. At each point we measure two velocity components and need to compute three components. Therefore, we require at least two symmetry points. Within the central diamond there are four equivalent points in the $z$ slice and there are extensions where three points are used as we lose the $x>0,y<0$ quadrant (in reality we use eight and six equivalent points as we assume symmetry $\pm z$ but these additional points do not give an independent view of the velocity field).The two regions correspond to accurate recovery of $(17,9,8)\mathrm{km\,s}^{-1}$ for the median absolute deviation. This is comparable to the input uncertainties of $20\,\mathrm{km\,s}^{-1}$. Outside these regions we primarily only have two observations (and only one for the four points at $(x,y)=(\pm 3.5,\pm0.6)\mathrm{kpc}$). Within these regions the median absolute deviations are $(60,32,11)\mathrm{km\,s}^{-1}$ although the deviation distribution has fat tails. We find that on average the recovery of the mean velocity field is unbiased to less than $1\,\mathrm{km\,s}^{-1}$.

\begin{figure}
    \centering
    \includegraphics[width=.85\columnwidth]{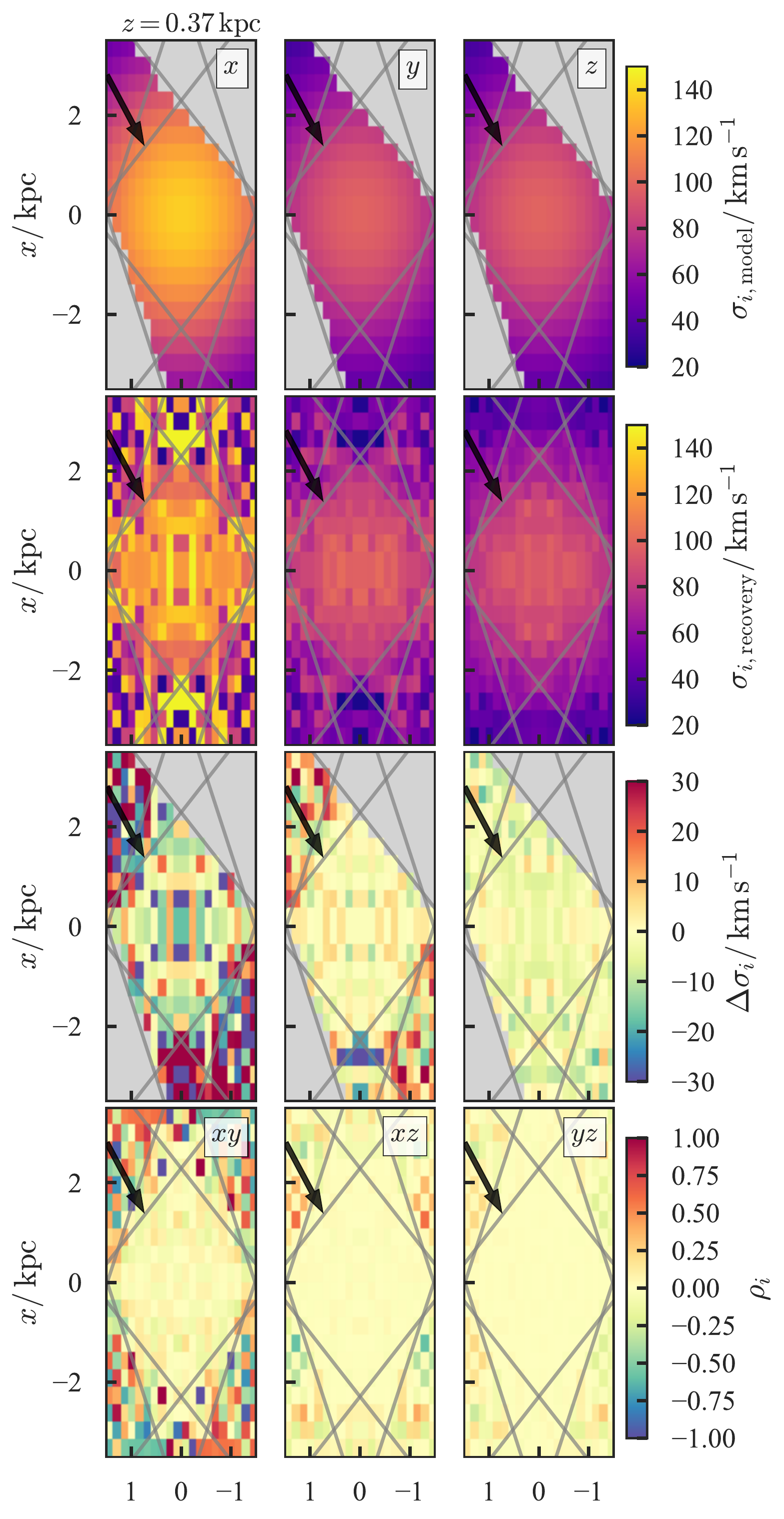}
    \includegraphics[width=.84\columnwidth]{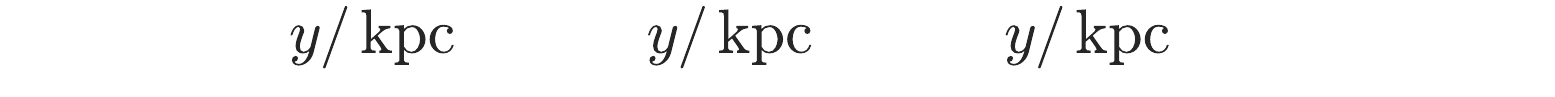}
    \caption{Recovery of the dispersion field from mock transverse velocity observations for a slice at $z=0.37\,\mathrm{kpc}$. Each column corresponds to a different velocity component in the bar frame ($x$, $y$, $z$). The first row is the mock `truth' $\sigma_i$, the second the recovery, the third the residuals and the fourth the recovery of the correlations (zero in the mock data and labelled by the inset). The grey lines show the selection volume limits reflected about the symmetry axes and the arrows show the observation direction.}
    \label{fig:disp_field}
\end{figure}

In Fig.~\ref{fig:disp_field} we show a slice through the velocity dispersion field. As with the mean velocities in Fig.~\ref{fig:mean_field} we observe structure associated with the selection volume. We measure three components of the velocity dispersion tensor which we use to find six unknown components. Therefore, we expect two independent observed symmetry points is sufficient for recovery of the field. In the regions where we have three or four observed symmetry points we well reproduce the structure of the dispersion field with uncertainties of $\Delta\sigma_i\approx(16,4,5)\,\mathrm{km\,s}^{-1}$ comparable to the input uncertainties of $10\,\mathrm{km\,s}^{-1}$. Our mock field has zero cross-terms everywhere. In the recovery we find the amplitude of the variation in $\rho_{ij}$ is $(0.1,0.02,0.01)$ for $(xy,xz,yz)$ with no significant bias. In the regions where we have fewer than three observed symmetry points the recovery of the dispersion field is poor, as expected. The median absolute deviations of the diagonal terms is $(54,18,5)\,\mathrm{km\,s}^{-1}$ but they are biased by $\sim20\,\mathrm{km\,s}^{-1}$ in $\sigma_x$ (recovery larger). Similarly the correlations are poorly reproduced with median absolute deviations of $(0.6,0.11,0.06)$ and biases of $\sim0.1$ in $\rho_{xy}$ (negligible bias in the other two components). We argued that two symmetry points was sufficient for recovery, which is not the case in practice. Two of our components inform us about the dispersion in the near vertical direction ($b$) whilst only one gives significant information about the in-plane field. Therefore, we expect that to compute three in-plane components from one component measured at each symmetry point we require at least three observed symmetry points. In conclusion, the recovery of the dispersion field in regions where we have fewer than three observed symmetry points should be treated with caution.

\bsp	
\label{lastpage}
\end{document}